\documentclass[aps,prb,twocolumn,showpacs,preprintnumbers,amsmath,amssymb]{revtex4}
\usepackage{dcolumn}% Align table columns on decimal point
\usepackage{bm}% bold math
\usepackage{amsmath}
\usepackage{feynmp}

\usepackage{graphicx}% Include figure files
\begin{document}

\title{Magnetoresistance when Spin Effects on Conduction are Weak}

% Use letters for affiliations, numbers to show equal authorship (if applicable) and to indicate the corresponding author
\author{Vincent Sacksteder IV}
\email{vincent@sacksteder.com}
\affiliation{Department of Physics and Astronomy, Rutgers University, NJ, USA}


\begin{abstract}
This paper considers certain materials, including topological insulators, where spin rotation symmetry is broken much more strongly than time reversal symmetry. When these materials are in the diffusive regime, i.e. when they have disorder that is strong enough to cause an electron to scatter many times while crossing a sample, electrons and holes move in pairs that have zero spin and are insensitive to spin  physics.  Working within this spinless scenario, we show that Fourier transforming the magnetoconductance with respect to external magnetic field obtains a curve describing the area distribution of loops traced by electrons and holes within the sample.  We present loop area distributions of Landau levels, weak (anti)localization, conduction governed by Levy flights, and linear-in-field resistance.  Of these four the last two  are new results.  Comparing  these distributions, we argue that the linear-in-field resistance seen in some topological insulators is caused by the same diffusive scattering that causes weak antilocalization.  The difference is that linear-in-field resistance materials retain  a level of quantum coherence that is usually seen only on the surface of 2-D wires or in ring geometries.  In an appendix we include  some speculative material about linear-in-temperature resistance.
\end{abstract}

%\dates{This manuscript was compiled on \today}
%\doi{\url{www.pnas.org/cgi/doi/10.1073/pnas.XXXXXXXXXX}}

%\begin{document}

% Optional adjustment to line up main text (after abstract) of first page with line numbers, when using both lineno and twocolumn options.
% You should only change this length when you've finalised the article contents.
%\verticaladjustment{-2pt}

\maketitle
%\thispagestyle{firststyle}
%\ifthenelse{\boolean{shortarticle}}{\ifthenelse{\boolean{singlecolumn}}{\abscontentformatted}{\abscontent}}{}

In atomic units $m_e = e = \hbar =  k_B = 1$ the magnetic field $B$ has  dimensions of inverse area, which is key to   understanding  the electrical conductance's dependence on $B$.  
 % If we pick c=1 instead of the a.u. definition c=1/alpha=137, then we end up having to choose (1/4 pi epsilon_0)=a_0 / alpha = 137 instead of the a.u. definition (1/4 pi epsilon_0) = a_0 = 1.  So the length scale is smaller  (and lengths are longer) but the Coulomb interaction constant is also bigger so it compensates to get the same energies.
Large fields probe  areas at the scale of the unit cell, and small changes in field probe  much larger areas.  At large fields one finds  Landau levels and Schubnikov-de Haas (SdH) oscillations, %which are used for measuring Fermi surfaces.  These
which  are visible in the longitudinal conductance
% Tony Leggett thought that this term longitudinal conductance needs clarification, but I forget why.
 $G_{xx}(B)$ as peaks at characteristic field strengths  determined by the Fermi energy $E_F$.  The Fermi surface's cross section, an inverse area, can be read off from the peak positions.    Landau levels and SdH oscillations are signals of ballistic physics, i.e. weak scattering, and are extinguished once scattering becomes too frequent. 

 At small fields one finds weak (anti) localization (WL/WAL), where  scattering and quantum interference  cooperate to cause $G_{xx}$ to  decrease or increase with field  depending on whether  spin rotation symmetry is broken faster or slower than time reversal symmetry.    %, or increase in the opposite case.   
WL/WAL contrasts strongly against Landau levels and SdH oscillations.  It is much more  sensitive to magnetic field, is visible at much smaller field strengths, and  varies smoothly  with field rather than exhibiting oscillations.  Moreover WL/WAL is caused by physics at  length scales much longer than the scattering length. It is only weakly sensitive to  the atomic unit cell and the Fermi wavelength.    

Recent years have exposed mysteries which lie well beyond traditional SdH and WL/WAL physics.  Many experiments report systems where the longitudinal resistance $R_{xx}$  increases linearly with $B$ and does not saturate. 
\cite{zhang2012magneto,PhysRevLett.106.156808,PhysRevLett.108.266806,PhysRevB.92.081306,kisslinger2015linear,PhysRevB.92.041203,PhysRevB.89.125202,xu2015quasi,PhysRevB.93.104502,PhysRevB.84.100508,bhoi2011quantum}
 This contrasts both with SdH oscillations which oscillate periodically in $1/B$ and with WL/WAL which is logarithmic.
% and typically much weaker than a polynomial.  
Explanations have been given for the cases of   ballistic conduction when the Fermi surface has a cusp, of a 3-D Dirac cone in the presence of a strong magnetic field, of a density gradient across the sample, and of classical transport with strong sample inhomogeneities. \cite{PhysRevLett.95.247201,PhysRevB.88.060412,PhysRevB.58.2788,PhysRevLett.117.256601,parish2003non}  However this kind of case by case treatment is not entirely satisfactory given the wide range of experimental realizations.  

The electrical conductance is determined by electrons and holes.  Magnetic field affects electrons and holes in two ways.  The first is  the minimal coupling  $\vec{p} - e \vec{R}$, where $\vec{R}$ is the gauge field determining the magnetic field by $\vec{B} = \vec{\nabla} \times \vec{R}$.  This coupling is sensitive only to a fermion's kinetics, not its spin.   In contrast, the $e \vec{\sigma} \cdot \vec{B}$ Zeeman term provides a very direct coupling between field and spin.  The Zeeman term is accompanied by similar terms which  are of higher order in momentum $\vec{p}$, electric field $\vec{E}$, and magnetic field.  \cite{winkler2003spin}

 In this paper we concern ourselves with materials where the Zeeman term's effect on conduction can be neglected, leaving in play only the minimal coupling.  In other words, we confine ourselves to systems where electronic conduction is mediated by spin singlets; we explore where this simplification can lead us.  Always working under the assumption that the Zeeman term can be neglected,  we show that  external magnetic field is conjugate to the area of the electron-hole loops inside a material.  In  non-interacting systems the conjugate area is that of a single loop, and in interacting systems it is the sum of the areas of all loops contributing to the conductance. 
  Therefore if one analyzes the magnetoconductance by taking the Fourier transform with respect to external magnetic field, one can extract detailed information about the length and area scales in play in electronic conduction within the material.  We apply this methodology to Landau levels, SdH oscillations, and weak antilocalization.  We go on to see what this methodology says about materials with linear-in-field magnetoresistance, if their linear resistance is caused by the minimal coupling not the Zeeman term.  Lastly we include some more speculative material concerning materials displaying linear-in-temperature resistance.

There exist many materials where electronic conduction is determined by spin singlets only; in these materials the spin triplet and the Zeeman effect have no influence. There are two ingredients for obtaining this scenario.  First, the material should be in the diffusive regime, where the scattering length is much shorter than the sample size.  Second,  spin rotation symmetry must be broken more strongly than than time reversal symmetry, as is the case with strong spin-orbit coupling.  

The first ingredient for spin-singlet conduction is diffusive conduction, with many scatterings as electrons and holes traverse a sample.  At lengths exceeding the scattering length $l$ the electronic wave-function $| \psi \rangle$'s phase is randomized, rendering single wave-functions unable to mediate conduction.  Therefore conduction over distances longer than $l$ is mediated only by pairing between $| \psi \rangle$ and its complex conjugate $\langle \psi |$, whose random phases cancel.  This pairing produces  the single-particle density matrix $\rho(x)$, which is the key mediator of conduction at distances longer than the scattering length.  This  result is  well known from the extensive work on sigma models and bosonization techniques in disordered systems. \cite{schafer1980disordered,efetov1999supersymmetry} %This point has been understood clearly for many years, as witnessed by the widespread use of sigma models and bosonization techniques for analyzing conduction. These approaches focus on the long-range physics and neglect physics at the scattering length, and find naturally that the appropriate variables are bosonic spins formed from $|\psi\rangle$  and $\langle \psi |$. 

$\rho$'s evolution is sensitive only to processes where  $| \psi \rangle$  and $ \langle \psi |$  undergo the same scattering events and follow the same path  between scatterings, because only these processes satisfy the requirement that the disorder-induced random phases  of $| \psi \rangle$  and $ \langle \psi |$  must cancel.   There are two such processes, the diffuson and the Cooperon.  In the diffuson  $| \psi \rangle$  and $ \langle \psi |$ follow the same sequence in the same order, producing purely classical electronic motion.    In contrast the Cooperon is a quantum interference process, distinct from  classical motion.   In the Cooperon $| \psi \rangle$  and $ \langle \psi |$ follow the same sequence but in reversed order, and their phases seem to be randomized with respect to each other. It is not  until their trajectories close a loop and come back to their origin that $| \psi \rangle$  and $ \langle \psi |$'s phases suddenly cancel and $\rho(x)$ undergoes a revival. Comparing the diffuson and the Cooperon, the diffuson is purely classical and depends weakly on magnetic field, while the Cooperon is purely quantum mechanical and is strongly dependent on even very weak magnetic fields.   

%It is the Cooperon's contribution to conduction which is responsible for WL/WAL.   
%  WL/WAL is caused by pairs of matching loops, one for  $| \psi \rangle$  and one for $ \langle \psi |$, following identical trajectories but in opposite directions.   

The main  effect of spin on disordered transport is that $\rho(x)$ contains both a spin singlet (charge) and a spin triplet (spin polarization).  Both the spin singlet and the spin triplet have characteristic relaxation times $\tau_s$ and $\tau_t$  that are determined by the material and by embedded impurities.  When the spin singlet relaxes more slowly than the spin triplet, i.e. when $\tau_s \gg \tau_t$, then the spin singlet contribution dominates and the spin triplet's contribution to conduction is neglegible.  This combination of lifetimes $\tau_s \gg \tau_t$ is found when spin rotation symmetry is broken more strongly than than time reversal symmetry.  This  is the case in materials with strong spin-orbit coupling, including topological insulators.  In particular,  when topological insulators are in the diffusive regime and their charge carriers are from  the Dirac cone not from other bands, then conduction is mediated only by spin singlets, not spin triplets.

%We have made one assumption throughout this discussion: that the phase factor associated with flux through electron and hole loops  is responsible for linear resistance, and not the Zeeman coupling between field and spin, which we have neglected.   As we have discussed earlier, disordered transport is mediated by the single particle density matrix $\rho(x)$.  

If the Cooperon increases the conductance when the magnetic field is zero, then its effect is called weak antilocalization (WAL).  If on the other hand the Cooperon decreases the conductance at zero field, then this is called weak localization (WL).     WAL results in the resistance increasing with field, while WL results in the resistance decreasing with field.  Whether weak localization is seen in a material, or instead weak antilocalization, is determined by whether the Cooperon  triplet is present.

Magnetic field is a perfect diagnostic tool for distinguishing between systems where the Cooperon spin triplet is absent and ones where it is present.  %At small fields in diffusive materials, the cause of magnetic field dependence is weak (anti)localization.  
When the Cooperon spin singlet relaxes more slowly than the spin triplet (as in topological insulators), i.e. when $\tau_s \gg \tau_t$,  the singlet produces weak antilocalization (WAL), which means that the resistance increases with field.  On the other hand, when time reversal symmetry is not more favored than spin rotation, and the  spin singlet and spin triplet relaxation time scales $\tau_s \sim \tau_t$ are similar, then the spin triplet dominates and the resistance decreases with field, which is called weak localization (WL).   \cite{hikami1980spin,PhysRevLett.48.1046}  In summary, WAL occurs when the spin triplet relaxes quickly  so that  conduction is mediated by charges which effectively have zero spin \cite{PhysRevB.83.241304}, while WL occurs when both the spin singlet and spin triplet  contribute to conduction.    

Experimentally, if the resistance increases with small fields in a diffusive material then only the Cooperon spin singlet is active in conduction, and not the Cooperon spin triplet.  If, on the other hand, the resistance decreases with small fields then both the spin singlet and the spin triplet are in play.

%If the spin singlet or spin triplet relaxation time $\tau_s$ and $\tau_t$ are short then these contributions are reduced.  In particular, , then the spin singlet contribution dominates and the net effect is a decrease in resistance, i.e. weak antilocalization (WAL).  

Many materials displaying linear magnetoresistance do conduct only via the Cooperon spin singlet, not the spin triplet.  Certainly topological insulators lie in this regime, as do Dirac semimetals with strong spin-orbit coupling.  \cite{zhang2012magneto,PhysRevLett.106.156808,PhysRevLett.108.266806,PhysRevB.92.081306}  Possibly other linear resistance materials, besides TIs and Dirac Weyl materials, may also have negligible spin triplet conduction.   Experimental observations of linear magnetoresistance universally find that resistance increases rather than decreases with field, as seen in WAL.  Observations of resistance increasing with field are consistent with a short spin triplet relaxation time $\tau_t$ and a significantly longer spin singlet lifetime $\tau_s$, and with dominance of spin singlet conduction.
 \subsection{Loops and Magnetic Field} This paper develops a geometrical analysis of magnetotransport in terms of the loops traced by electrons and holes as they move through real space.  In spin singlet materials where only the minimal coupling and not the Zeeman term couples magnetic field to electrons, electron-hole loops are controlled via the magnetic flux through them and by the Aharonov-Bohm effect.  Because magnetic flux is equal to the product of field with loop area, field is conjugate to area.  We will develop this fact into a methodology for analyzing experimental measurements of the conductance.
 
   The starting point of our analysis is Feynman's formulation of quantum mechanics as a sum over paths. \cite{feynman1965quantum}
  In this formulation an electron follows not just one path, but instead many paths which fully explore the sample in which the electron is moving.  All of these many paths are summed to determine the evolution of $|\psi \rangle $, the electronic state.     In this paper our focus is not on the state $|\psi \rangle$, but on the many paths that contribute to $|\psi \rangle$.  The two pictures are mathematically equivalent: starting from a knowledge of   paths one can build  the electronic propagator which controls the evolution of $|\psi \rangle$, and conversely  $|\psi \rangle$'s evolution may  be systematically decomposed into the paths which determine it.

  To be clear, we are not discussing the semiclassical paths of an electron's average motion.  For instance, in a magnetic field an electron's average position executes well-defined circular loops around the axis of the magnetic field, and one can measure  this cyclotron motion. This is not the sort of path we are talking about.  Instead we are talking about the infinitely many quantum mechanical paths, tracing many complex trajectories and fully exploring the sample, which sum up to produce the average cyclotron motion.
   
  The paths traced by electrons never begin or end in isolation, since in this event electronic charge would not be conserved, i.e.   charge would be generated or destroyed. Instead,    the creation or annihilation of an electron is always accompanied by the creation or annihilation of an accompanying hole with opposite charge, and by this mechanism charge is conserved.  The two paths of an electron and of its corresponding hole, followed from their origin together to their final disappearance together, form a loop.  
  
  It is important to be clear that the loops under discussion here are traced by bare electrons and holes, not by quasiparticles.   The  requirement to move in loops  is an immediate consequence of charge conservation and gauge invariance, and does not require  that interactions are weak, the existence of a  Fermi liquid, or well-defined quasiparticles. Nor does it not assume any other sort of collective many-body behavior.   The loop requirement applies independently of magnetic fields, scattering, and  interactions.   Analysis focused on electron-hole loops is therefore a very powerful  way of understanding the mechanisms responsible for electronic transport.

It is also important to be clear that the loops discussed here move through real space, and occur even when disorder is very strong.  Therefore they are not the same loops used in the analysis of Berry phases and topological invariants, which are usually calculated in momentum space and assume a well defined band structure.  Moreover the loops discussed here concern the bare electron and hole carriers within a material; the U(1) symmetry which supports them comes from electromagnetism, not from effective interactions that occur in spin liquids and other interacting systems.  % not sigma models.   not quantum computation, 

 Experimental data on a sample's dependence on external magnetic field gives direct access  to the areas of the electron and hole loops within the sample.  Each  loop couples to field via a phase which is proportional to the magnetic flux through the loop, and this phase is equal to  the product of the loop area and the magnetic field strength.  We neglect the Zeeman term, assuming that only the spin singlet contributes to conduction. Within the phase factor magnetic field has a conjugate relationship to area, with a proportionality constant that is determined by fundamental constants and can not be renormalized.    Therefore measurements of a sample's dependence on external magnetic field  give information about the area of the  loops traced by electrons within the sample. More precisely, the area that one learns about is the sum of the areas of all loops that contribute to the conductance.  %, and provide a window onto the physical processes and length scales at work in transport.  
 In a non-interacting system this simplifies to the area of a single loop.
  The methodology of using magnetic fields to measure loops, which we explain further below, does not assume quasiparticle or Fermi liquid physics. It is a simple consequence of charge conservation. In interacting systems
 
In this article we apply this methodology focused on electron-hole loops and their areas to electrical conduction.  We obtain results about the electron and hole loops responsible for Landau levels, SdH oscillations, WL/WAL, and linear magnetoresistance.   Our rigor here is limited only by our omission of the Zeeman effect, which amounts to an assumption that only the spin singlet contributes to conduction.  Where this assumption is valid,  there is a firm link between experimental observables and loop areas.  Details of the material's Hamiltonian, eigenstates, interactions, Fermi surfaces, or disorder do not affect this link, and  are  discussed very little in this paper. Instead we focus on reasoning about the electron and hole loops that are responsible for conduction, and on drawing conclusions about which physical mechanisms are responsible for producing those loops.

\subsection{What We Can Learn  From Loop Areas} 
Our methodology is to start with the conductance's dependence on external magnetic field, and then Fourier transform with respect to magnetic field.  The resulting quantity depends on magnetic field's conjugate quantity, the total area of all loops that contribute to the conductance.   We call this quantity the loop area distribution.  The final step is to examine the weights of this distribution, i.e. the weight corresponding to a particular value of the total loop area, and allow these weights to inform us about what electrons and holes are doing inside the material.

One reference point for comparison is the Landau levels and SdH oscillations, which are phenomena that do not rely on scattering.  The loop area distribution for these effects oscillates periodically with loop area, with a period of oscillation that is  determined by the Fermi surface.   It has both positive and negative signs. Both the presence of a particular area scale, and oscillations between negative and positive signs (i.e. ringing), are hallmarks of ballistic conduction.

A second  reference point is weak (anti)localization (WL/WAL).  Here the loop area distribution has a single sign, and does not show any concentration near any particular area scale (i.e. it is scale-free). \cite{hikami1980spin,efetov1999supersymmetry,rammer2018quantum}  The weights of the loops responsible for the WL/WAL signal  decrease gradually and smoothly with area, and the decay follows a $1/A$ (where $A$ is area) power law up to loop areas which are far larger than that determined  by the Fermi surface.    The very smooth distribution of loop areas and the very large area scales  are caused by stochastic motion of the charge carriers, and are  unmistakeable hallmarks of scattering-driven (diffusive) conduction.   
  
   We next compare the loop area distribution responsible for linear magnetoresistance to that of standard WL/WAL, Landau levels and SdH oscillations.  We are making  the assumption (explicit throughout this paper) that  in the linear resistance materials of interest neither spin triplets nor the Zeeman term affect electronic conduction.  This is certainly a reasonable assumption for those topological insulators which manifest linear magnetoresistance.   
   
   The theoretical predictions for Landau levels, SdH oscillations, and WL/WAL are all based on  2-D electron gas models with no interactions.  Therefore the loop area distribution is that of single electron-hole loops.  In principle a linear resistance material may be strongly interacting. In this  case the loop area distribution would concern the total area of all loops contributing to conduction, and comparison to non-interacting loop area distributions like those of Landau levels or WL/WAL may be misleading. Fortunately this is not a concern in   most topological insulators, which  are only weakly interacting.
      
   We find that the loop area distribution corresponding to linear  magnetoresistance is very like the loop area distribution of WL/WAL.  There are no changes in sign, there is no favored length scale, the area distribution decreases smoothly as area increases, and very large area scales are seen.  This is  evidence that scattering and stochastic motion are key to linear magnetoresistance. 
 % and that both WL/WAL and linear magnetoresistance are caused by pairs of matching loops, one for  $| \psi \rangle$  and one for $ \langle \psi |$, following identical trajectories but in opposite directions.   These paired loops are the Cooperon, whose  distribution of loop areas is scale free and smooth and extends to very large areas, as seen both in standard WL/WAL and in linear magnetoresistance.
 This is also evidence that linear magnetoresistance is caused by Cooperons, i.e. paired particles and holes, rather than by any species of single-particle motion.  The phase of single electrons and holes oscillates vary rapidly at the Fermi wavelength, which is the ultimate reason why the loop area distribution of Landau levels exhibits an oscillating sign.  In contrast the phase of a Cooperon varies at the much longer localization length, resulting in a very smooth single-sign loop area distribution.

  Despite these similarities, there is one major difference between WL/WAL and linear magnetoresistance: the latter has a much longer tail than the former; the loop area distribution of linear magnetoresistance decays like $\ln A$ while WL/WAL decays like $1/A$.  In other words, linear magnetoresistance has many more large loops, and much larger loops.   This reweighting towards large loop areas is the opposite of what would be obtained by transitioning from 2-D conduction to 3-D conduction. It is impossible to explain if Cooperons move diffusively in 2-D, whether with ordinary random walks, or with Levy flights.
  
  The extraordinarily long and fat tail seen in linear magnetoresistance shares something in common with Landau levels, whose loop area distribution is a simple cosine.  In the absence of scattering and at zero temperature, this cosine extends to infinitely large areas, signifying that all sizes of electron-hole loops  are in play.  The reason is not, of course, that the electrons are exploring an infinite 2-D plane. Instead the repeated oscillations of the loop area distribution signify that if an electron can complete one cyclotron orbit, then it can complete any number of cyclotron orbits.  The electron's ability to retrace its path around a loop is a manifestation  of quantum coherence.  If finite temperature disrupts quantum coherence, its  effect is to cut off loops with large areas.
  
  In two dimensional samples the Cooperons that produce WL/WAL do not repeat their loops; they are effectively random walkers moving according to purely classical dynamics, without quantum coherence.  The reason is that scattering randomizes momentum, and therefore even when a random-walking Cooperon returns to its starting point, it starts off from there in a new direction. This is a peculiarity of 2-D planar materials.  When Cooperons are confined to the 2-D surface of a wire, or move one dimensionally around a circle, then they do manifest quantum coherence and repeat their loops.  
  
  Linear magnetoresistance involves much larger loop areas than can be explained via Cooperons undergoing ordinary random walks. We therefore conclude that the Cooperons in linear magnetoresistance materials retain some degree of quantum coherence, allowing them to sometimes repeat their loops.  In other words,  linear magnetoresistance is caused by a combination of quantum coherence and scattering - a combination that is difficult to account for in planar 2-D materials.   Further discussion of this topic is in Section \ref{QuantumCoherence}.

Section \ref{GeometricAnalysis}  will begin with charge conservation and show that it implies that magnetic field is conjugate to area. This  allows one to start with experimental measurements of the conductance's dependence on magnetic field and then obtain  information about the areas of electron and hole loops. Next section  \ref{AreaDistributions} applies this methodology to calculate and compare the loop area distributions of  Landau levels, SdH oscillations, WL/WAL, linear magnetoresistance, and Levy flights.  Section \ref{QuantumCoherence}  discusses linear magnetoresistance and possible links to quantum coherence, and section \ref{Final} wraps up with a few final thoughts.  Appendix \ref{AlternateRegularizations} discusses several ways of regularizing the Fourier transform from $1/A$ to the logarithm and vice versa, and Appendix \ref{LinearInTemperature} gives some more speculative comments on materials with linear-in-temperature resistance, such as bad metals.

 % Keimer and Zaanen: % In a normal metal, unless the metal melts first, the re- sistivity saturates at high temperatures when the mean free path, l, becomes of the order of the electron de Broglie wavelength. % This quote assumes that there is no quantum interference and no anderson localization.

\section{Geometric analysis of electron and hole loops\label{GeometricAnalysis}}

Charge conservation  is a fundamental feature of nature which has been subjected to intense scrutiny and has been found to be obeyed to extreme precision.   Two strategies are available to  engineer a theory to correspond to this experimental reality. One strategy is to select a set of quantum mechanical states available to the system, and a Hamiltonian which describes  the system's ground state and dynamics.  In the process of building the states and the Hamiltonian, one enforces a constraint: all states must have the same total charge, and the Hamiltonian may not change that charge.   This approach is very powerful, and has shown extreme versatility in building models of the essential physics of many systems.  These include simplified models where the  electron's position is restricted to an integer number of discrete values and therefore electron paths are series of discrete jumps.  However this approach has the weakness that charge conservation is essentially imposed by fiat, and that its fundamental connection to the geometry of electron and hole loops is lost.  

To capture the geometry of electrons moving in the real world, it is necessary to start from the symmetries of the real world.  These include continuous translation, i.e. the electrons' ability to move smoothly along continuous paths instead of jumping from position to position, and Poincare symmetry, the ability to smoothly change direction in the four dimensional manifold of space and time.   The class of theories which unite these symmetries to quantum mechanics, relativistic quantum field theory, is our starting point for understanding the connection between charge conservation and geometry.

An early and essential result of relativistic quantum field theory was the Dirac equation, which predicted the existence of holes.  Soon afterwards quantum electrodynamics was formulated, in which the principle of gauge invariance is key to guaranteeing charge conservation.  Later the constraints of internal consistency (renormalizability, unitarity, etc.) and of  experimental data resulted in the standard model, in which gauge invariance is elevated to a guiding principal that is key to all electronic interactions.  The standard model incorporates quantum electrodynamics, and preserves its treatment of charge conservation of electrons and holes, which is based on gauge invariance.

Gauge invariant theories are built in two steps.  The first step is adding a gauge field: a $U(1)$  phase $\phi(\vec{x})$ which depends on position.  This phase modifies the momentum operator which describes translational motion, so that as an electron moves along its path, its phase is multiplied by the difference between the phases  $\exp(\imath\phi)$ at the electron's final position and its initial position. The second and final step toward building a gauge invariant theory is the requirement that the theory be invariant under arbitrary changes of the gauge field.   This is essentially a mathematical shorthand for imposing a restriction on the structure of every physical observable.    The  only way to build an observable $\mathcal{O}$ which is independent of $\phi(\vec{x})$  is by requiring each electron  contributing to the observable to  return to its starting position,  forming closed loops.  Because the final and initial positions are the same, the difference in phase $\exp(\imath\phi)$ between start and end is exactly zero, and therefore observables built from closed loops do not depend on  $\exp(\imath\phi)$; they are gauge invariant.  The actual shape and size of the loops, and also the number of loops,  is completely unrestricted by gauge invariance; the emphasis is on their closed nature.

In relativistic quantum field theory the electron paths move through space-time, and the requirement to return to the origin does not mean  only that an electron must  reverse its motion in space.  Each electron is  also required to reverse its direction in time, so it can eventually return to its starting time.  From one point of view this reversal of direction is simply a manner of pulling a loop's path back to its origin, but from another point of view the reversal of time-direction is the meeting (creation or annihilation) of a particle-hole pair.   In other words, one can adopt a vocabulary that avoids discussing electrons moving backwards-in-time, but only at the cost of saying that for every electron [moving in forward in time] there is also a hole.  The hole is thought of as moving forward in time, but it has exactly the same mathematical and physical consequences as an electron moving backward in time.  The net effect is that instead of talking about electrons tracing loops in space-time, one requires that electrons and holes are always produced and destroyed in pairs.  If one takes the hole and electron paths together one finds that they always add up to loops in space-time.  This loop structure is an immediate and unavoidable consequence of combining gauge invariance with special relativity.

Gauge invariance ensures charge conservation in a way that is so powerful that it seems almost trivial, as follows. First it requires that that observables be made out of space-time loops. Secondly we divide up each space-time loop into segments that move forward in time [electrons], and segments that move backward in time [holes].  Thirdly we assign charge $-e$ to electrons and charge $+e$ to holes.  Fourthly we note that the loop structure implies that  electrons are always created or destroyed with corresponding holes, in pairs.  We thus arrive at charge conservation:  this structure of pair-wise creation/destruction manifestly conserves total charge.  

\subsection{Coupling Magnetic and Electric Fields to Loops}
In relativistic quantum field theory the effect of magnetic and electric fields on electrons is mediated by the minimal coupling: the momentum operator  $p^\mu$ is replaced by $ p^\mu - e R^\mu$, where the gauge potential $R^\mu = \partial^\mu \phi$ is  the first derivative of the gauge phase $\phi(\vec{x})$.  The electric and magnetic fields $\vec{E},\vec{B}$ are encoded in $R^\mu = (\phi/c, \vec{R})$ via $\vec{E} = -\vec{\nabla}\phi - \partial \vec{R}/\partial t$ and $\vec{B} = \vec{\nabla} \times \vec{R}$. This minimal coupling is required to construct a gauge invariant continuum theory - it is necessary to ensure  that when an electron moves from a starting position to a final position, it is multiplied by the difference in phases $\exp(\imath\phi)$ at the two positions.

The minimal coupling implies that as an electron or hole traces its path, the only effect of electric and magnetic fields is to introduce a multiplicative factor, the Wilson loop $W_\gamma = \exp(\imath  \oint_\gamma ds_\mu   R^\mu)$.  \cite{PhysRevD.10.2445,PhysRev.80.440}  The line integral $\oint_\gamma ds_\mu$ is taken along the electron/hole path, denoted by $\gamma$.  It is a loop integral because electron and hole paths   are required to form loops.

When several electron-hole loops occur the line integrals of each loop add up, and the multiplicative factors $\exp(\imath  \oint_\gamma ds_\mu   R^\mu)$ multiply each other.    Therefore we shift  the notation from  $\gamma$ which specifies a single loop to  $\Gamma$, which  specifies the paths traced by a set of one or more electron and hole loops.  The total effect of all the loops together is to multiply by  $W_\Gamma = \exp(\imath  \oint_\Gamma ds_\mu   R^\mu)$, where the line integral $\oint_\Gamma ds_\mu $ is the sum of each of the loop integrals $\oint_\gamma ds_\mu   R^\mu$ in $\Gamma$, summing over the one or more loops specified by $\Gamma$.   Gauge invariant observables are built from sums in which each term of the sum has a particular set of electron-hole loops specified by $\Gamma$ and is weighted by the corresponding Wilson loop $W_\Gamma$.

By Stokes' theorem the phase $ \oint_\Gamma ds_\mu   R^\mu \equiv  \Phi_\Gamma$ in the  Wilson loop is equal to  the electro-magnetic flux through the one or more loops specified by $\Gamma$. In other words, the line integral $\oint_\Gamma {ds} $ over loops $\Gamma$ is equivalent to a flux integral  $\int_\epsilon {dS}$ over  the surfaces $\epsilon$ bounded by the loops.  Mathematically $ \oint_\Gamma ds_\mu   R^\mu = \Phi_\Gamma =  \int_\epsilon {dS_{\mu \nu} F^{\mu \nu}}$.   Here $F^{\mu \nu} = \partial^\mu R^\nu -  \partial^\nu R^\mu$ is the electromagnetic field tensor whose components are equal to the electric and magnetic fields, and $dS_{\mu \nu}$ is the infinitesimal surface tensor for a surface embedded in four-dimensional space-time.  When the surfaces  $\epsilon$ are all fixed to a particular time $t$  this flux  simplifies to the magnetic flux $\Phi_\Gamma =  \int_\epsilon {d\vec{S}} \cdot \vec{B}$, where now ${d\vec{S}}$ is the surface infinitesimal  in three-dimensional space and $\vec{B}$ is the magnetic field.

 In atomic units where $e = \hbar = 1$ the phase $\Phi_\Gamma$ controlling the Wilson loop  $W_\Gamma = \exp(\imath  \Phi_\Gamma)$ is equal to the electromagnetic flux through all the loops specified by $\Gamma$. Because both the phase and the flux are strictly the argument of an exponential, they must be unit-free.  In turn the units of electromagnetic field $F^{\mu \nu}$ must be the inverse of the units of the area infinitesimal $dS_{\mu \nu} $.  This is the fundamental reason why the electromagnetic field tensor $F^{\mu \nu}$ , and magnetic field $\vec{B}$ in particular, have units of inverse area.

 %: the only meaning of a magnetic field is its action on a closed electron loop, which is equivalent to the flux through any surface bounded by that loop.

We formalize these ideas using the   loop representation of observables. \cite{ashtekar1992loop}  The previous discussion can be summarized by the statement that  every physical observable $\mathcal{O}_i(\vec{B})$  must be written as
\begin{eqnarray}
\label{FirstSumOverLoops}
\mathcal{O}_i(\vec{B}) = \sum_\Gamma \langle \mathcal{O}_i | \Gamma \rangle \exp(\imath  \Phi_\Gamma)
\end{eqnarray}
 where $\Gamma$ specifies the spatial trajectories of all the electron-hole loops, and $ \sum_\Gamma $ is a sum over all the ways these electron-hole loops can be arranged - i.e. both their number and their individual trajectories.   $\mathcal{O}_i$, is the observable,  and $ \langle \mathcal{O}_i | \Gamma \rangle $ is the weight with which a particular Wilson loop   $\Gamma$ contributes to  $\mathcal{O}_i$.   This weight   is determined by both the observable and by the kinetics and interactions of the electrons and holes tracing the loops specified by $\Gamma$.

\subsection{The non-relativistic limit}
  In relativistic quantum field theory the electron-hole pair follows a path which curves through space time and finally returns to its origin.  The  energy of the electron and the energy of the hole, taken separately, vary throughout their motion around the loop.  
  
  The non-relativistic limit is taken by assuming that the energies of the electron and hole are almost constant and remain very near their rest energy/mass.    Obviously this requires a strict conceptual division between electron and hole.   In this limit the Dirac equation, describing both electron and hole, simplifies to the Pauli equation which describes either the electron or the hole.  
  
  Because of the non-relativistic limit's special treatment of the time dimension, electric field drops out of the minimal coupling and takes a separate role in the Pauli equation, and is joined by the $e \vec{\sigma} \cdot \vec{B}$ Zeeman term.  The remaining minimal coupling is  $\vec{p} - e \vec{R}$, where $\vec{R}$ is the gauge field determining the magnetic field by $\vec{B} = \vec{\nabla} \times \vec{R}$.  The minimal coupling  produces Wilson loops $\exp(\imath \Phi_\Gamma)$ where the flux $\Phi_\Gamma =  \int_\epsilon {d\vec{S}} \cdot \vec{B}$ depends only on magnetic field not electric field.  
  
In the non-relativistic limit magnetic field affects electronic motion not only through Wilson loops but also  via the Zeeman term.  As we have argued earlier, in a class of systems which includes topological insulators in the diffusive regime, the spin triplet relaxation time  $\tau_t$ is short compared to the spin singlet relaxation time $\tau_s$.  Therefore  spin and the Zeeman term do not contribute to electronic transport over long distances \cite{PhysRevB.83.241304}, and observables measuring transport depend on magnetic field only via Wilson loops, not via the Zeeman term.

 \subsection{External Fields and the Loop Area Distribution}

   %  \textbf{ Distinguish between external field H and the total field B.  For example in a superconductor B=0. It is only in an almost free electron picture  i.e. quasiparticle/Fermi liquid picture that we can consider the tracks of single particles and the magnetic field running through them. }

 We specialize to the case of a uniform magnetic field $\vec{H}$ originating externally to the sample, which we will call $\vec{B}^{ext} \equiv \vec{H}$.  We consider the effect of $ \vec{B}^{ext} $ on the one or more electron-hole loops specified by $\Gamma$.  Each electron-hole loop experiences not only the external field $\vec{B}^{ext} $, but also magnetic field generated by  the electron-hole loops within the sample.  The total field, including both external field and the field generated by the sample, is called $\vec{B}$.  Total field $\vec{B}$ is related to  external field $\vec{B}^{ext} $ via $\vec{B} = \mu_0 \vec{B}^{ext}  + \mu_0 \vec{M}$, where $\vec{M}$ is the magnetization. $\mu_0$ is the vacuum permeability, which we will drop for simplicity.  In diamagnetic materials, including superconductors, $\vec{B}$ is smaller than $\vec{B}^{ext}$, while in paramagnetic materials $\vec{B}$ is larger than $\vec{B}^{ext}$.   
 We separate the magnetic flux through a Wilson loop into  two terms: a flux $\Phi_\Gamma^{int}$  caused by the field generated by the sample,  and a flux $\Phi_\Gamma^{ext}= \int_\epsilon {d\vec{S}} \cdot \vec{B}^{ext} $ from external field.   Therefore in an external magnetic field observables can be written as
 \begin{eqnarray}
\label{SecondSumOverLoops}
\mathcal{O}_i(\vec{B}^{ext}) = \sum_\Gamma \langle \mathcal{O}_i | \Gamma \rangle \exp(\imath  \Phi_\Gamma^{ext})
\end{eqnarray}
The phase  $\Phi_\Gamma^{int}$ from field generated within the sample is now packaged inside the weight  $ \langle \mathcal{O}_i | \Gamma \rangle$.   The phase $\exp(\imath  \Phi_\Gamma^{ext})$ depends only on external field, not on internally-generated field as well.

 Because the external field $\vec{B}^{ext} $ has uniform direction and magnitude, it is useful to assign to each Wilson loop a  vector-valued cross-section $\vec{A}_\Gamma = \int_\epsilon d\vec{S}$, which is a vector generalization of the Wilson loop's surface area.  This area vector depends only on paths traced by the loop $\Gamma$, not on the shape of the surfaces $\epsilon$ which are bordered by $\Gamma$. The magnetic flux through $\Gamma$  then simplifies to the dot product of the area $\vec{A}$ with the external field $\vec{B}^{ext} $; $\Phi_\Gamma = \vec{A}_\Gamma \cdot \vec{B}^{ext}$.  This leads  to our central result:
     \begin{eqnarray}
         \label{AtoBTransform}
\mathcal{O}_i(\vec{B}^{ext}) &=&  \int {d\vec{A}} \;   \mathcal{O}_i(\vec{A}) \exp(\imath  \vec{A} \cdot \vec{B}^{ext}), 
\\ \nonumber 
  \mathcal{O}_i(\vec{A}) &\equiv  & \sum_\Gamma \langle \mathcal{O}_i | \Gamma \rangle \delta(\vec{A}_\Gamma  - \vec{A})
\end{eqnarray}
In other words, every physical observable can be resolved into contributions corresponding to specific areas $\vec{A}$, each with weight $\mathcal{O}_i(\vec{A})$:
%, and this weight can be obtained experimentally by first measuring the observable's dependence on magnetic field and then performing a Fourier transform:
     \begin{eqnarray}
     \label{BtoATransform}
\mathcal{O}_i(\vec{A}) &=& \int \frac{d\vec{B}^{ext}}{(2 \pi)^3} \;   \mathcal{O}_i(\vec{B}^{ext}) \exp(-\imath  \vec{A} \cdot \vec{B}^{ext}) 
\end{eqnarray}
The area $\vec{A}$ is the sum of the areas of all the electron-hole loops that contribute to the observable.  

In summary, equations \ref{AtoBTransform} and \ref{BtoATransform}  state that measurements of an observable $\mathcal{O}_i(\vec{B}^{ext})$ 's  dependence on external magnetic field give direct and rigorous information about the sum of the areas traced out by electron loops within a sample, simply by performing a Fourier transform.   The resulting loop area  distribution  $\mathcal{O}_i(\vec{A})$ completely describes how different loop areas contribute.  %In particular, the external field $\vec{B}^{ext}$ is conjugate to a specific area $\vec{A}$.  %As long as only the spin, and there is no possibility that any process could renormalize this relation.  

From a theorist's point of view, the problem of how to perform a regularized sum over loops $\sum_\Gamma $, i.e. how to control both long distance and short distance contributions to the sum, remains in general unsolved.  This makes evaluation of the sum over loops $\sum_\Gamma $ in equations \ref{FirstSumOverLoops} and \ref{AtoBTransform} a difficult task.  It is therefore both remarkable and encouraging that  we can measure the sum experimentally and directly using  equation \ref{BtoATransform}; experimental measurements of the loop area distribution $\mathcal{O}_i(\vec{A})$ can lead the way in guiding theorists to correct procedures for performing the sum over loops.

%\subsection{Mathematical Expressions of Gauge Invariance, Solid State}
%strongly interacting lattice gauge theory where loops traverse the lattice and in perturbation theory where electrons and holes always contribute via loops of one or more concatenated Green's functions.

\section{Comparison of Loop Area Distributions\label{AreaDistributions}}
Throughout the remainder of this paper $\vec{B} = \vec{B}^{ext} = \mu_0 \vec{H}$ means the external magnetic field, and $B = |\vec{B}|$ is its magnitude.

\subsection{Landau levels and SdH Oscillations} To illustrate the transformation from the magnetoconductance to the loop area distribution, consider the case where the dominant energy scale is the Fermi level $E_F$, in which case a delta function - the first Landau level - will be found in $G_{xx}(B)$ at $B^{-1}= \frac{1}{2}  A_F$. Here $2 \pi / A_F \propto 2 \pi E_F $ is the cross-section of the Fermi surface. \cite{onsager1952interpretation}  The loop area distribution is therefore a cosine $G_{xx}(A) \propto \cos(A / \frac{1}{2} A_F)$, which means that loops with every area contribute to the conductance and that there is ringing corresponding to the characteristic Fermi area $A_F$.  \footnote{A slightly different result $ \cos(A / N A^F)$ obtains for the $N$-th Landau level of Dirac fermions. \cite{PhysRevLett.82.2147}}  %  clarify why there are loops at B=0 and why they depend on E_F. "why should we interpret the motion of particles in the absence of a field in terms of loops at all?" 

The most interesting aspect of this result is that quantum coherence causes  the first Landau level to be repeated  exactly at   characteristic areas $(N+1/2)A_F$,  with the $N$-th level's height proportional to $N$.  The repetitions are manifested as a hierarchy of additional Landau level delta functions in $G_{xx}(B)$, as illustrated in Figure 1a.     Figure 1b shows the loop area distributions $G_{xx}(A)$ of the Landau levels.  The $N$-th Landau level is a cosine    $\cos(A / (N+\frac{1}{2} ) A_F)$ with period equal to $(N+1/2)\;2\pi$  times the characteristic area $A_F$.   In mathematical terminology, the $N$-th Landau level is simply the $N$-th harmonic of the lowest Landau level.  Speaking more plainly, quantum coherence ensures that if an electron can complete a loop once, then it can repeat that same loop any number of times. 

       \begin{figure}%[tbhp]
\centering
\includegraphics[width=1\linewidth,clip,angle=0]{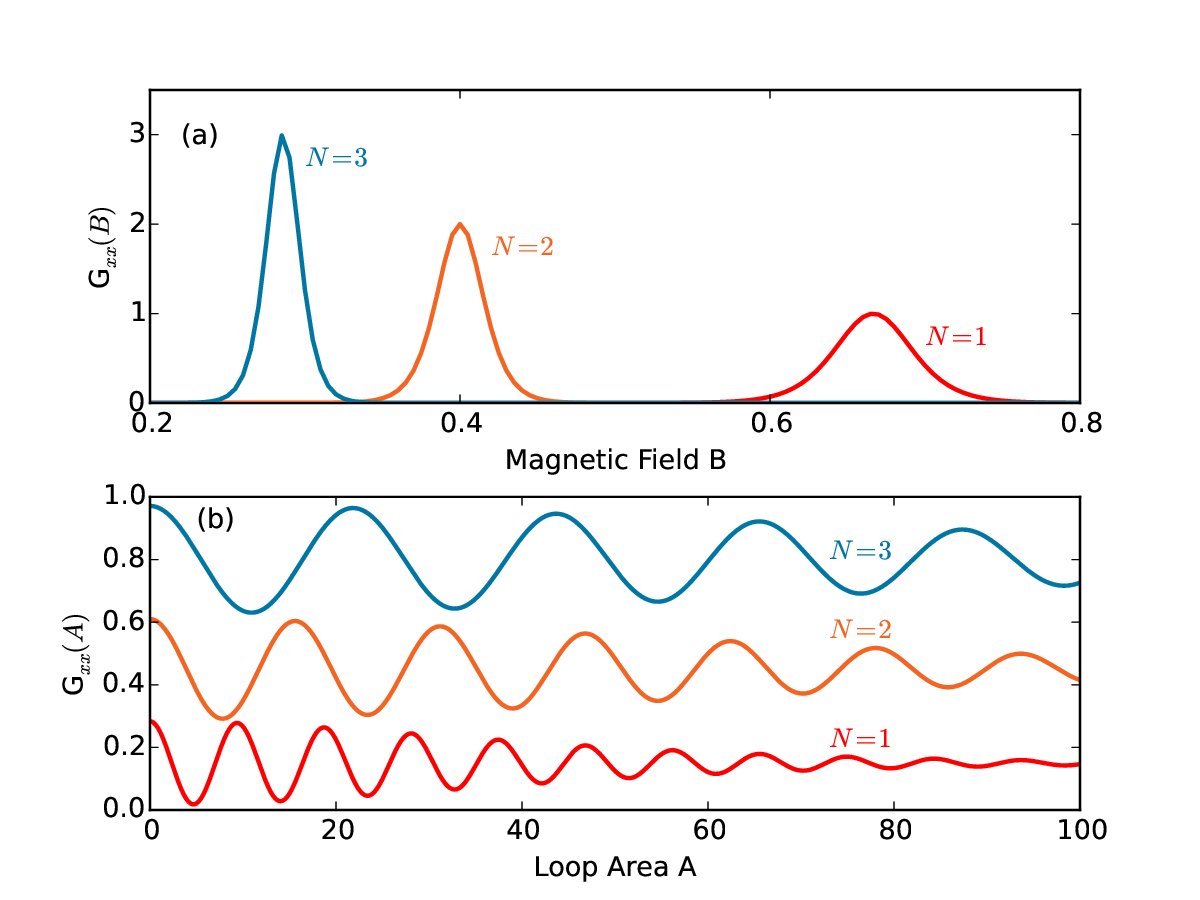}
\caption{ (Color online.)  Quantum coherence and the hierarchy of Landau levels. Panel (a) shows the magnetoconductance $G_{xx}(B)$  of the $N=1,2,3$ Landau levels and panel (b) shows their loop area distributions $G_{xx}(A)$, which have been vertically shifted for clarity. \cite{laikhtman1994quasiclassical}  Quantum coherence is responsible for producing higher harmonics of the lowest Landau level, causing higher Landau levels at multiples $(N+\frac{1}{2} ) A_F$ of the characteristic area $A_F = E_F^{-1}$.  Decoherence, i.e. loss of quantum coherence, is caused by either the scattering energy scale $\Gamma$ or the temperature  $T$. Decoherence widens peaks in $G_{xx}(B)$, suppresses $G_{xx}(A)$ at large $A$, and completely suppresses oscillations when the field $B$ is as small as $\Gamma, T$.   $E_F=1,\;\Gamma = 0,\;T=0.05$.   
}
\label{AreaHierarchy}
\end{figure}

The only limit to these repetitions (harmonics) is  decoherence caused either by the scattering energy scale $\Gamma$ or by the temperature $T$, which causes a  % is this a power law decay?  I wrote that down, presumably because of my numerical work on the Fourier transform of the SdH/Landau level spectrum, but I didn't keep any notes.
decay in  the loop area distribution.  \cite{laikhtman1994quasiclassical}  Decoherence broadens  the Landau levels into peaks with width $\Gamma, T$, which eventually merge into   SdH oscillations.  These  oscillations have equal height instead of the height proportional to $N$ seen in Landau levels.  They
%, and its area distribution narrows into a peak centered around $ A_F$.  
% \footnote{In Dirac materials the peak is shifted to $A_F/2$.
 decay exponentially when the value of $B = N^{-1} A_F^{-1}$ descends to $\Gamma, T$, and their loop area distribution also decays exponentially at areas larger than $\Gamma^{-1},T^{-1}$. \cite{laikhtman1994quasiclassical} \footnote{To verify that loop area distribution for Landau levels and SdH oscillations decays exponentially, we numerically Fourier transformed the conductance formulas in Ref. \cite{laikhtman1994quasiclassical}, because analytical formulas for the Fourier transform of this signal are not available. }
\subsection{Weak Antilocalization}
%, which is caused by Wilson loops $\Gamma$ composed of pairs of electron and hole loops, called Cooperons, which have charge $q_C = 2e$.  These pairs are much more  robust against disorder than single electrons, whose phase  is randomized after each scattering.  In contrast, the phases of a Cooperon's  electron  and  hole  cancel each other because the two follow the same sequence of scattering events in reversed order.     % This electron-hole pair,  taken as a whole, is called a Cooperon. 
%Two electron loops which follow the same path but in time reversed order will have the same amplitude and phase at $B=0$, regardless of scattering.   

We now investigate 2-D WL/WAL, whose magnetoconductance $G_{xx}^{WAL}(B)$ has been studied extensively.   We will show  that the WL/WAL loop area  distribution  varies inversely with area i.e. it obeys  $G_{xx}^{WAL}(A) \propto A^{-1}$.     At magnetic fields less than the inverse cutoff  $q_C B  < 1/L_{max}^2$ the conductance $G_{xx}^{WAL}(B)$  is controlled by the cutoff: it quickly transitions to zero, and this transition is quadratic in $B$ because time reversal symmetry requires that $G_{xx}$ be even under $B\rightarrow -B$.     At larger fields $q_C B > 1/L_{max}^2$ the magnetoconductance of a 2-D spin singlet is  a logarithm $G_{xx}^{WAL}(B) \propto -\ln B$.  This logarithmic form, which will be our focus here, is robust and ubiquitous.  If spin or orbital degrees of freedom are in play then the logarithmic conductance seen from the Cooperon spin singlet generalizes simply to a sum of logarithms %with different weights
 corresponding  to  the Cooperon spin singlet,  the Cooperon spin triplet, and possibly other Cooperon sectors associated with valley degrees of freedom etc.   The logarithmic conductance  is seen for both  Dirac and $p^2/2m$ dispersions, for both lattice and continuum models, for both short and long range scattering, and for a wide variety of quasi-2-D geometries. \cite{hikami1980spin,al1981magnetoresistance,dugaev1984magnetoresistance,PhysRevB.39.11280,raichev2000weak,PhysRevLett.89.206601,PhysRevB.86.035422,PhysRevB.38.3232,PhysRevB.88.041307,PhysRevB.90.235148,PhysRevB.90.045408,PhysRevLett.99.146806,PhysRevLett.99.106801,PhysRevB.77.081410} \footnote{At very strong magnetic fields, when the magnetic length $\sqrt{\hbar c / eB}$ is compared with the mean free path, there is a further transition to $G_{xx} \propto 1/\sqrt{B}$ behavior.  \cite{PhysRevB.56.9910,gasparyan1985field,dyakonov1994magnetoconductance,cassam1994two}  }
The logarithm is found in all 2-D systems that are in the diffusive regime, i.e. all 2-D systems whose size is larger than the momentum relaxation length caused by scattering.

The reason that the same form of $G_{xx}^{WAL}(B)$ is universally found in diffusive 2-D systems  is that at distances exceeding the momentum relaxation length the Cooperon loses all information about momentum and kinetics, and therefore its trajectory is simply a  2-D random walk.    When spin or orbital degrees of freedom are in play the Cooperon carries several species of random walkers (e.g. a spin singlet and  a spin triplet), but each of the species is still performing a simple random walk.  The WL/WAL magnetoconductance signal is determined  by Cooperon trajectories which return to their origin, forming loops.  Quantum coherent repeats of loops do not contribute  because the Cooperon's momentum is random and therefore even when a Cooperon returns to its origin it does not have the same momentum that it started with.   Since the trajectories of these Cooperon loops are simply random walks, the magnetoconductance  signal has the same form in all 2-D diffusive systems.   

The fact that the universal form of the magnetoconductance $G_{xx}^{WAL}(B) \propto \ln B$ is logarithmic was first determined by Ref. \cite{hikami1980spin}. % using a technique where the Cooperon was described using a diffusion kernel appropriate for describing random walks.   
 Ref. \cite{hikami1980spin} starts with a description of how scattering changes the motion of the electronic wave-function $|\psi \rangle$ and its complex conjugate $\langle \psi |$, and then uses this information to derive the behavior of the Cooperon, which is a pairing of   $|\psi \rangle$ with $\langle \psi |$.  It is found that the spin singlet component of the Cooperon is simply a random walker and that every random walk returning to the origin contributes with equal weight to the WAL conductance $G_{xx}^{WAL}(B)$. Mathematically this is expressed by writing the Cooperon Green's function as the inverse of a Laplacian operator which expresses how individual steps in the walk are taken.  To derive  from disordered scattering the Cooperon and its random walk, Ref. \cite{hikami1980spin} uses the standard field theory of disordered scattering, bosonization, and sigma models. \cite{schafer1980disordered,efetov1999supersymmetry}  In this respect it is generally imitated by all other theoretical (non-computational) treatments of  Cooperons and WL/WAL.  Always the result is that WL/WAL is mediated by random walking Cooperons.
  
   The WAL conductance is  proportional to the probability that the Cooperon will return to its origin; mathematically it is the diagonal $\vec{x} =\acute{\vec{x}}$ matrix element of the Cooperon Green's function.  At zero field  $G_{xx}^{WAL}(B=0)$ is simply the probability that random walkers will return to their origin, while non-zero field causes each walker trajectory to be reweighted by a Wilson loop phase factor.   Calculation of the Cooperon spin triplet component is almost identical to that of the singlet - the main difference is that the triplet contribution to $G_{xx}$ is multiplied by $-1$, producing weak localization instead of weak antilocalization.

 We emphasize the fact established in  Ref. \cite{hikami1980spin}  that all trajectories contributing to the Cooperon's spin singlet  component contribute with the same weight, with no phase factor from kinetics, scattering, potentials, or time evolution.  Likewise, all trajectories contributing to the Cooperon's triplet component contribute with the same weight.  In other words, although the Cooperon describes a special kind of quantum interference process that is  exhibited by electrons, the actual behavior of a Cooperon is no different than that of a purely classical random walker.  No phase information enters in  because a Cooperon is a pairing of  $|\psi \rangle$ with $\langle \psi |$, and their phases cancel each other.   Therefore the Cooperon evolves as a classical random walker, and calculations of $G_{xx}^{WAL}(B)$  look purely classical. All Cooperon trajectories (of the singlet, or of the triplet) contribute with the same sign and weight.  One very important consequence is that the loop area distribution $G_{xx}^{WAL}(A)$ of the Cooperon singlet (or triplet) trajectories has only one sign, like a purely classical probability distribution.  Calculation of WL/WAL is isomorphic to calculating return probabilities of a purely classical walker.  
      
     This picture of a purely classical Cooperon is modified in samples with sizes approaching the localization length, where the conductance is reduced to the order of the universal conductance quantum $G_0$.  At this long length scale processes coupling two or more Cooperons become important, as do processes coupling Cooperons with diffusons.  However the systems which show the \textit{weak} localization and antilocalization which is our focus here lie far from the localized limit, so that the WL/WAL  contribution remains at the level of simple random walks without multi-Cooperon processes.
 %The main  effect of spin on disordered transport is that $\rho(x)$ contains both a spin singlet (charge) and a spin triplet (spin polarization).  In materials with strong spin-orbit coupling or strong spin-dependent scattering the spin polarization relaxes very quickly and only the spin singlet contributes to electronic conduction.  In other words, when the spin relaxation length is short conduction is mediated by charges which effectively have zero spin.  In this case the Zeeman coupling's  effect on electronic transport is negligible. 
 %The fact that resistance increases rather than decreases with field is also strong evidence that the spin triplet is negligible.    The singlet contribution to WL/WAL  decreases the resistance and therefore when magnetic field disrupts WL/WAL the resistance increases; this is called weak antilocalization.  In contrast, the spin triplet contribution has the opposite sign causing weak localization, so if the spin relaxation length is long magnetic field causes the resistance to decrease.  Since experimental observations of linear magnetoresistance universally find that resistance increases rather than decreases with field, we conclude that the spin relaxation length is short, and that the Zeeman effect is not relevant to conduction.

 We demonstrate here that the Fourier transform of the logarithmic WL/WAL conductance  $G_{xx}(B) \propto \ln B $ varies inversely with area, i.e. that the loop area  distribution obeys  $G_{xx}(A) \propto A^{-1}$.  In other words the Fourier transform of $-\ln B$ is $A^{-1}$ and, vice versa, the inverse Fourier transform of $A^{-1}$ is  $-\ln B$. The main difficulty is controlling the ultraviolet and infrared cutoffs, so we perform the demonstration three ways.  First, we use purely dimensional analysis: $G_{xx}(A) = (2 \pi)^{-1} \int {dB} \exp(-\imath A B) \ln B$. Therefore $G_{xx}(A)$ must carry units of inverse area; it must be equal to $A^{-1}$ times some function that depends only on dimensionless quantities. 
 
 Second, we use specific functional forms for $G_{xx}(A)$ and carry out analytically the inverse Fourier transform from $G_{xx}(A)$ to $G_{xx}(B)$.  For instance, we consider hard cutoffs at $A_0$ and $A_{max}$, with $G_{xx}^{WAL}(A) = A^{-1}$ for $A_0 < A < A_{max}$.  With these cutoffs  we arrive at $ G_{xx}^{WAL} (B) \approx R_0 -   \ln |q_C B | + (q_C B A_0)^2/4$ for $1/A_{max} < q_C B < 1/A_0$, where $q_C$ is the Cooperon charge, $R_0 = -\gamma_E -\ln |A_0| $,  and $\gamma_E$ is the Euler-Mascheroni constant. Here $R_0$ should be adjusted to zero, as a regularization compensating for the hard cutoffs, to replicate the well-known fact that  the WAL (Cooperon spin singlet) contribution to the  conductance is always positive for all values of $B$, and that the triplet WL contribution is always negative.  \cite{hikami1980spin}  The appendix imposes several other cutoffs on $G_{xx}^{WAL}(A) = A^{-1}$, in each case performs the Fourier transform exactly,  and  in each case obtains logarithmic forms for $G_{xx}^{WAL}(B ) = \int dA \, \exp(\imath A B) A^{-1} \propto -\ln B$.

    Lastly, Ref. \cite{hikami1980spin} and many subsequent works have established that the magnetoconductance $G_{xx}^{WAL}(B )$ of a 2-D random walking Cooperon is logarithmic. Therefore we can show that the Fourier transform of $-\ln B$ is $A^{-1}$ simply by showing that the loop area distribution of 2-D random walkers is  $G_{xx}^{WAL}(A )\propto A^{-1}$.   The reasoning is like a syllogism: random walkers have a logarithmic magnetoconductance, and random walkers have an $A^{-1}$ loop area distribution.  The loop area distribution is the Fourier transform of the magnetoconductance; therefore the Fourier transform of $-\ln B$ is $A^{-1}$.
    
    Figure 2a  summarizes the results of extensive  numerical Monte Carlo simulations of random walks, all of which verify that the loop area distribution of random walks which return to their origin scales with $A^{-1}$.  We use an infinite 2-D square lattice, and define the lattice spacing equal to one.  The walkers take steps which are chosen from a Gaussian distribution with mean step length $l=4\sqrt{2},8\sqrt{2}$, and an upper  bound (IR cutoff) to the loop areas is supplied by limiting the walk length to $5 \times 10^5$.   We count the walkers which return to their origin, compute the area of the loop traversed by each walker, and histogram the area distribution $G_{xx}(A)$.   Panel (a) shows $G_{xx}(A)$, multiplied by $A$.   The plateaus of constant  $G_{xx}(A) \times A $ extending over several orders of magnitude verify that 2-D random walks follow an $A^{-1}$ loop area distribution. We have verified that similar results are found with other cutoffs and with other probability distributions governing the step lengths, and a similar numerical Monte Carlo simulation of 2-D random walks has previously verified the $A^{-1}$ loop area distribution. \cite{PhysRevB.61.13164}   These numerical demonstrations of the $A^{-1}$ loop area law again verify that  the Fourier transform of $-\ln B$ is  $G_{xx}(A) \propto A^{-1}$, and vice versa.

%The results in all of these cases are so similar
%, and the logarithmic and quadratic forms are so featureless, 
%that it is experimentally very difficult to extract from  WL  data any information beyond $L_{max}$ and  $\alpha$, the number of conducting channels.
% two or more coupled planes, when the 2-D sample is given a finite depth with or without scattering in the bulk, and when the electronic motion is on the 2-D surface of a 3-D TI wire or a nanotube.  
% cite Yongqing's papers.

%The reason why almost no information can be extracted from the HLN formula turns out to be that the formula has almost no content: the UV physics such as Landau levels and SdH oscillations has been removed, leaving room for  only two stylized facts.  The first is  the quadratic form at small $B$ - an infrared cutoff - and its form dictated by time reversal symmetry.  The second piece of information in the HLN formula is the logarithm.   

\begin{figure}%[tbhp]
\centering
\includegraphics[width=1\linewidth,clip,angle=0]{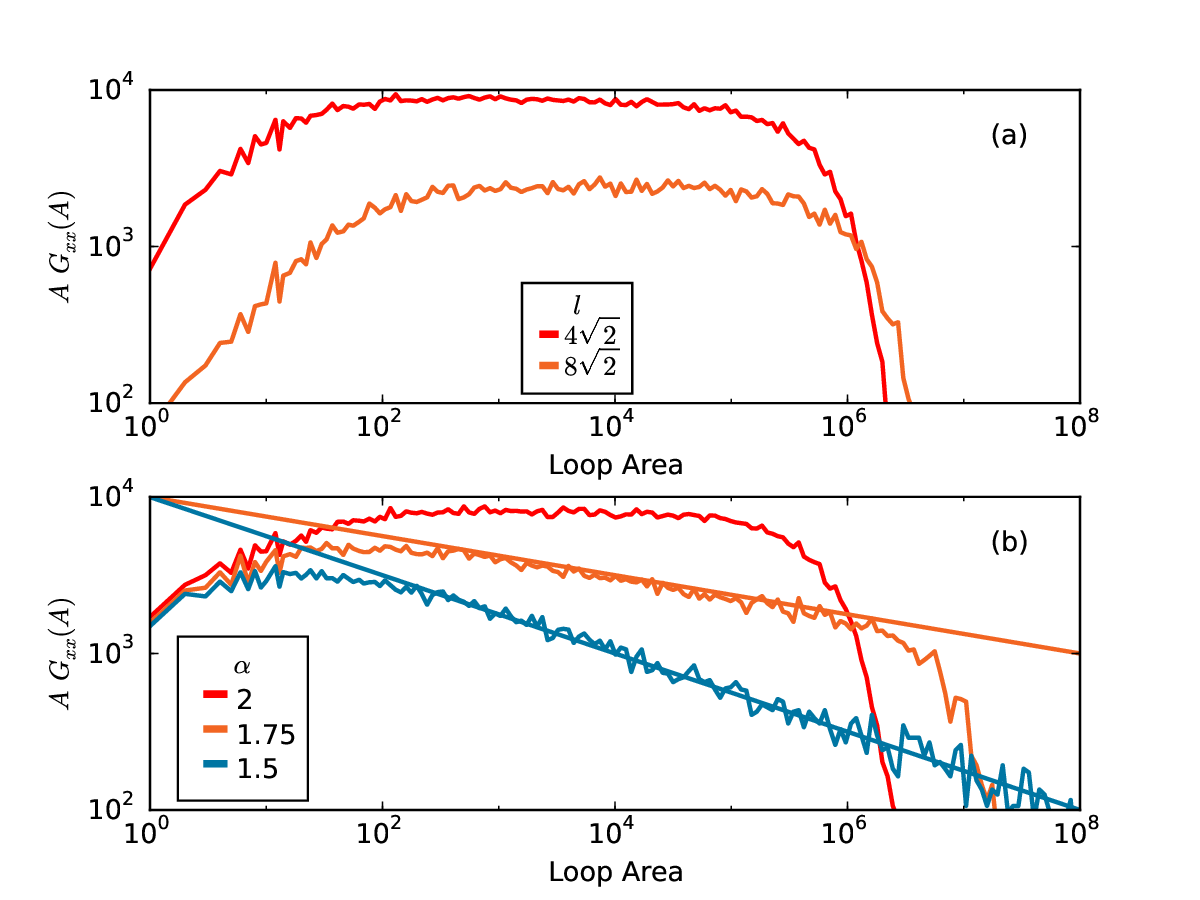}
\caption{ (Color online.)   The Cooperon's loop area distribution $G_{xx}(A)$ obtained by Monte Carlo simulations.   Panel (a) shows $G_{xx}(A)$, multiplied by $A$, of standard 2-D random walks with Gaussian-distributed steps having  average length $l=4\sqrt{2},8\sqrt{2}$.  The plateaus extending over several orders of magnitude verify that 2-D random walks follow an $A^{-1}$ loop area distribution.  The IR cutoff  is caused by limiting the walk length to $5 \times 10^5$.  Panel (b) shows that Levy flights have a loop area distribution which decays faster than $A^{-1}$.  $\alpha$ is the stability parameter of the Levy alpha-stable distribution which controls the step lengths.  $\alpha=2$ produces  Gaussian-distributed steps, and when $\alpha$ is decreased below $2$ the distribution develops a heavy tail of very long steps. The straight lines are for $10^4 \times A^{-0.125}$ and $10^4 \times A^{-0.25}$, and their agreement with the Monte Carlo data shows that $G_{xx}(A)  \propto A^{-2 + \alpha/2}$.  $l=4\sqrt{2}$ and the IR cutoff is $5 \times 10^5$, $1.5 \times 10^6$, and $5 \times 10^6$ for the $\alpha=2,\,1.75,\,1.5$ results.
}
\label{RandomWalks}
\end{figure}

% the WL logarithm is nothing more a statement about the areas of loops formed by random walks. Specifically, if one generates 2-D random walks, and selects from this ensemble walks that return to their origin, and measures the area $A$ of the loop traced by each random walk, then the probability of loops with area $A$ is $P(A) = 1/A \propto \sigma_{xx}(A)$. 

 \subsection{Linear magnetoresistance} This result can be reversed and applied to the case of linear magnetoresistance, where  $G_{xx}(B) \propto B^{-1}$.  We established above that the inverse Fourier transform of $A^{-1}$ is $-\ln B$; interchanging $A$ with $B$ shows that the loop area distribution corresponding to linear magnetoresistance is  logarithmic, $G_{xx}(A) \propto -\ln A $.   This profile, whose broad features are general for any roughly linear resistance, is remarkably different from  the $A^{-1}$ distribution that governs loops generated by random walks.  It has fewer small loops and  many more large loops;  a small head and a very  long tail.  We emphasize that the logarithmic loop area distribution is a rigorous result, as long as only spin singlets contribute to conduction and the Zeeman effect can be neglected.  If this assumption holds, then linear magnetoresistance  implies a logarithmic loop area distribution,   no matter what  physical mechanism is responsible for producing the loops. 
  % .   It shows that electron loops with very long length scales are required to obtain a linear magnetoresistance.

\subsection{Levy Flights and 3-D }
  Since linear magnetoresistance corresponds to a $\ln A$ distribution that decays very slowly with loop area, it would be of interest to determine how such a long tail can be produced.  However the $1/A$ decay seen in WL/WAL seems to be the slowest decay and the biggest tail that can be produced by purely stochastic random walks moving on a 2-D lattice. 
  %It is difficult to explain the logarithmic area distribution's  long tail using  completely random walks. 
  Figure 2b  shows the loop distribution that is obtained if  the Cooperon does not follow a diffusive path where the step length distribution has a finite mean, variance, etc. Instead  the Cooperon in Figure 2b follows a Levy flight where step lengths follow a distribution  including both short and  very long steps.  The result is  a steeper decay than $A^{-1}$,  because   long steps tend to decrease the probability that the walker's path will complete a loop.    Here  we again use a 2-D square lattice, choose random steps from a probability distribution, and impose an IR cutoff $l$
   %$l = 5 \times 10^5, \; 1.5 \times 10^6, \; 5 \times 10^6$ 
   on the maximum path length.  
   The only difference from Figure 2a is that instead of choosing steps from a Gaussian probability distribution to produce a random walk, we choose steps from the Levy alpha-stable distribution, producing Levy flights where most steps are short but some steps are very long.   The Levy alpha-stable distribution is controlled by the Levy distribution stability parameter $\alpha$ which we give values $\alpha=2,\,1.75,\,1.5$.  When $\alpha = 2$ it reduces to the Gaussian distribution, and when $\alpha$ is decreased below $2$ the distribution develops a fat tail of very long steps. 
   Panel 2b plots  $G_{xx}(A) \times A$.  At $\alpha = 2$ we see again the plateau signaling the $1/A$ area decay law of random walks.  At $\alpha = 1.75,\,1.5$ the loop area distribution  decays faster than $A^{-1}$, tilting the plateau.     We superimpose straight lines for $10^4 \times A^{-0.125}$ and $10^4 \times A^{-0.25}$, and their agreement with the $\alpha = 1.75,\,1.5$  Monte Carlo data shows that $G_{xx}(A)  \propto A^{-2 + \alpha/2}$.  In summary, Levy flights steepen the area law decay, taking it further from the very slow $\ln A$ decay necessary to produce linear magnetoresistance.   
   
   Increasing the dimensionality above two dimensions also produces a steeper decay, $A^{-3/2}$ in 3-D. \cite{baxter1989fitting}  This is the opposite of what is wanted, so  one might like to instead decrease the dimensionality below two dimensions.  However in one dimension all loops have area identically equal to zero.    Therefore neither changing the step distribution nor changing the dimensionality can reproduce the $\ln A$ tail seen in linear magnetoresistance systems.
 % Other ways of producing a long tail: interactions killing off the small $A$ regime?  ballistic physics? interesting rules about backscattering during Levy flights?
 
 We note that in very strong magnetic fields where the magnetic length $l_H= \sqrt{\hbar c / eB}$ is smaller than  the mean free path  a $\sqrt{A}$ area law is obtained at  areas smaller than $l_H^2$. \cite{PhysRevB.56.9910,gasparyan1985field,dyakonov1994magnetoconductance,cassam1994two} However this produces a $1/\sqrt{B}$ magnetoresistance, not a linear signal, and  is strictly a large-field behavior, unlike experiments where  linear magnetoresistance starts at small fields.

  % Single Landau level: \cos(A/A_F), \sum_n \cos(A/n A_F), both positive and negative
  % SdH: \delta(A - A_F)??? or \sum_n \delta(A - n A_F)??? or something else with peaks?, only positive?
  % Cooperon Diffusion, Levy flight, 3-D: 1/A or steeper decay, positive
  % Linear resistance: \ln A
  % Superconductivity: Sin(A B_max) / A, i.e. flat and positive at small enough A and 1/A times ringing with negatives at larger A.

%  More broadly, a quasilinear resistance implies that the loop distribution $\sigma_{xx}(A)$ has a very long tail, decreasing very slowly with $A$, so that very big loops contribute much more strongly than they would if conduction were a random walk with uniform step size.  The question is what conduction process will  obtain such long tails in the loop area distribution.  

\section{Linear Magnetoresistance and Quantum Coherence\label{QuantumCoherence}}  

 Linear magnetoresistance's  $-\ln A$ loop area distribution shows no characteristic scale, decreases uniformly and  smoothly,  and  never changes sign.  These facts are strong evidence that  linear magnetoresistance is caused by Cooperons, i.e. by paired particles and holes scattering together, rather than by any species of single-particle motion.

The loop area distribution of  linear magnetoresistance exhibits a long fat logarithmic tail that can not be explained by random walks or by Levy flights.  Therefore we propose that the Cooperons responsible for linear magnetoresistance do follow random walks, but with the special feature that they maintain quantum coherence, allowing them to repeat their loops many times, in the same way that Landau levels are repeated in a hierarchy at $(N+\frac{1}{2})A_F$. 
% In this case $\sigma_{xx}(A)$ has a contribution from a loop with  area $a$, from the loop's second repetition with area $2a$, its third repetition with area $3a$, and so on.  

 As a simple example, consider the case of 2-D random walks with hard cutoffs at $A_0$ and $A_{max}$ and  $G^1_{xx}(B, A_0, A_{max}) = \int_{A_0}^{A_{max}} {dA} \cos(q_C B A) / A $.  The $1/A$ in this expression is the loop area distribution of standard random walks, and the cosine and integral perform the Fourier transformation from area to magnetic field. $q_C$ is the Cooperon charge. At field strengths between the UV cutoff $1/A_0$ and the IR cutoff $1/A_{max}$ this magnetoconductance $G^1_{xx}(B, A_0, A_{max}) $ follows the logarithmic form of standard WL/WAL.  
 
Now take that expression, and allow loops to repeat up to $N$ times, with a weight proportional to $1/N$.  This  procedure depletes the loop distribution's head, because if a loop area distribution has a UV cutoff at $A_0$, then its  $N=2$ first repetition will have a higher UV cutoff at $2 A_0$, its second repetition will have its cutoff at $3 A_0$, etc. Therefore between $A_0$ and $2 A_0$ the  total loop area distribution will have a contribution from  only the base $N=1$ loop distribution,  while at larger areas higher and higher $N$ will contribute.  The net effect is to  enlarge the tail of the loop area distribution and shrink its head.

Performing the sum over loop repetitions obtains   $G_{xx}(B) = \gamma_E + \sum_{n=1}^N n^{-1} G^1_{xx}(B, n A_0, A_{max})$.  We have added a regularization constant $\gamma_E$ to compensate for cutoff effects.    Figure 3a compares this loop area distribution to the diffusive $1/A$ profile produced by random walks without coherence.  The new loop area distribution has the required  small head and long tail; it is a logarithm.  Figure 3b  shows its Fourier transform, the  magnetoresistance. It is linear, i.e. proportional to $R_{xx}(B) - R_{xx}(B=0) \approx \pi q_C  B A_0 $, in the range $1/A_{max} < q_C B <  0.2 / A_0$. 
%This result is remarkable both for its clear linearity and for the excellent agreement of its slope $ \pi q_C A_0 $ with experimental results on cuprates and pnictides, which we will discuss later. 

 \begin{figure}%[tbhp]
\centering
\includegraphics[width=1\linewidth,clip,angle=0]{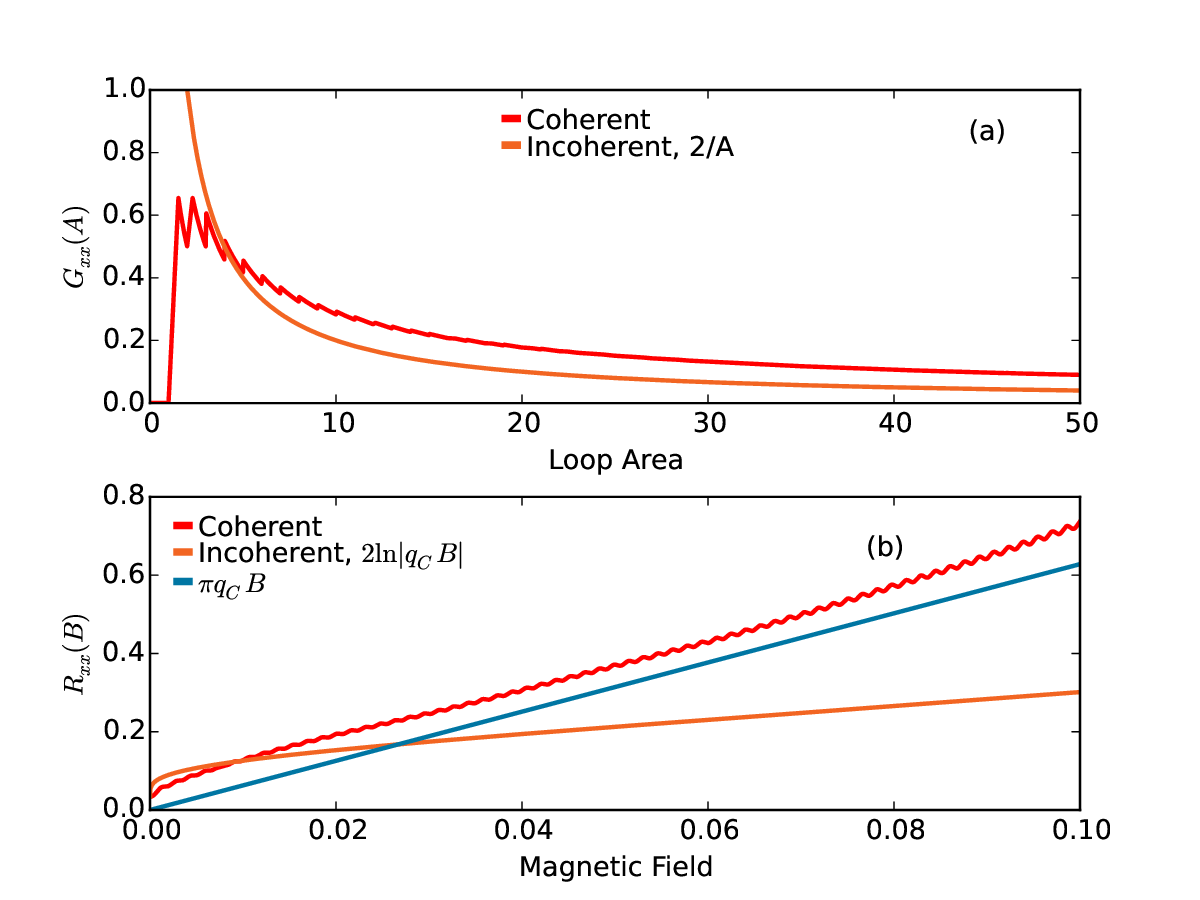}
\caption{ (Color online.)  Linear magnetoresistance from quantum coherence.  The red lines show  the results when quantum coherence allows loops to repeat up to $N=100$ times. They can be compared to the orange lines which show standard behavior without quantum coherence.   Panel (a) shows the loop area distribution $G_{xx}(A)$, where the quantum coherent distribution (red line) clearly has a small head and long tail as compared to the standard distribution (orange line).  Panel (b) shows the resistance $R_{xx}(B)$.  
%The blue line shows a perfect logarithmic distribution $G_{xx}(A) = \ln |A/80| / 2\pi$. 
  The quantum coherent result (red line) compares nicely to the blue straight line with slope $\pi q_C$. $A_0=1$, and the  sharpness of  the hard IR cutoff at $A_{max} = 2000$ causes the ripple in the quantum coherent result.   The orange line  gives a standard logarithmic curve, $R_{xx}(B) = (0.1 - 2 \ln |q_C B|)^{-1}$. 
}
\label{LinearResistance}
\end{figure}

This simple example is a proof of principle about how to obtain linear magnetoresistance. We chose an abrupt IR cutoff at  $A_{max}$ and an abrupt UV cutoff at $A_0$, but the true forms of these cutoffs will be material dependent. These cutoffs can have visible effects.   In Fig. 3b superimposed on the linear signal one sees a small amplitude fast ripple, which is ringing caused by the  abrupt IR cutoff.  A similar ripple has been seen in Ref. \cite{zhang2012magneto}, signaling that a sharp IR cutoff was present in that experiment.  However in most experimental data on linear resistance a ripple is not visible, indicating either  a smooth IR cutoff or that magnetic field was not sampled at fine enough resolution. At large fields the form of the UV cutoff also manifests in the signal's behavior.  Our sharp UV cutoff  causes  strong ringing at large fields  $q_C B > 0.2 / A_0$ [not shown in Fig. 3b] which overwhelms the linear magnetoresistance signal, but with a suitably chosen UV cutoff the linear growth will extend  to larger $B$.

 We note that linear magnetoresistance does not require the IR cutoff $A_{max}$, i.e. the area scale where phase coherence is suppressed, to be very large.  For instance,  a coherence area of $(100 \,\AA)^2$ would allow linear magnetoresistance to extend down to $\approx 0.3 $T.  Moreover, scatterers within a sample must not fluctuate or move more quickly than the time required for a Cooperon to complete its loop. However this time scale is extremely short.  Using $T^{-1} = 4A$ from Appendix \ref{LinearInTemperature}  and $\tau = \hbar / k_B T$, we find that $\tau$  is three picoseconds for loops of area $A = (100 \, \AA)^2$.
  % 2 * 2 *  ( 100^2 Angstrom^2 ) = 2 *  2 * ( 100^2 Angstrom^2 ) * (a_0/0.529 Angstrom)^2 * (Hartree * a_0^2)^{-1}  * (Hartree * s * 2.419 * 10^-17) = 2 * 1.729 * 10^-12 s
% As a consequence a diffuse (non-momentum-conserving) component should be seen in the single particle density of states and in excitation spectra such as ARPES, when analyzed as  functions of momentum. % \textbf{This is in fact true } in the cuprates and pnictides.  
%   We have suggested that the bad metals' linear resistance is caused by interaction-induced scattering.   In principle the scattering could be caused by anything that fluctuates at a time scale longer than the time $\tau $ required to go around a loop.  

% This cutoff should be visible as deviations from linearity when. what is $0.2  / q_c  A_0$ in Tesla? $2.35 \times 10^5 T = a_0^{-2}, 3.158 \times 10^5 K = a_0^{-2}$. Maybe there should be a factor of 2 or 1/2 when converting to temperature. The optimal hole concentration in ybco is around 1/(2000 a_0^2), so the magnetic field where deviations from linearity are found should be around 1/(20000 a_0^2) or 10T, and the corresponding temperature should be 15 K.  So really we need to do better, maybe 20/q_c A_0 would be good.
We emphasize that a power law area distribution can always be given a depleted head and long tail by invoking  quantum coherence and allowing loop repetitions. The form of the UV and IR cutoffs, the relative weight of each repetition, and even the base loop area distribution $G^1_{xx}(A)$ may be specific to the scattering  source and to the mechanism responsible for Cooperon coherence. There are only two fundamental requirements: a scattering process producing an area distribution with broad support across a range of areas, and quantum coherence allowing Cooperon loops to repeat in a hierarchy like that of Landau levels.  

The real difficulty here is that ordinarily  Cooperons in 2-D planar materials do not display loop repetitions; they do not retrace their path two or more times.  The fundamental reason  is that in 2-D (or 3-D, 4-D, etc.) when a Cooperon returns to its starting point it does so with a momentum that is different from its original momentum.  Therefore on the second time around the Cooperon leaves the origin moving in a direction that is different from its first time around.  We now turn to the question of how 2-D Cooperons could retain a degree of quantum coherence.

There are two exceptional cases which prove that under the right circumstances Cooperons are able  to maintain quantum coherence; they are able to show quantum  loop repetitions similar to the higher harmonics seen in Landau levels and SdH oscillations.  The most notable exception is 1-D systems with static disorder.  In 1-D the Cooperon's particle and hole are constrained to follow the same path, and the static disorder means that their energies never change, and neither does the magnitude of their momenta.  Therefore when a Cooperon returns to its starting point it can easily repeat the same exact loop that it has just ended, starting with a momentum that has the same magnitude and direction as before.  As a result Cooperon-mediated quantum interference dominates the behavior of 1-D and quasi 1-D systems: once the system's length $L$ increases enough so that the usual  classical $1/L$ Ohm's law decay brings the conductance down to one conductance quantum $G_0$, quantum interference takes over. This can take two forms. In most systems  the conductance decay transitions from the $1/L$ Ohm's law to the exponential decay characteristic of Anderson localization. \cite{ishii1973localization,simon1989trace,RevModPhys.69.731} However in topologically protected systems the conductance plateaus at a constant value of $G_0$, manifesting a perfectly conducting channel.   Both cases manifest complete dominance of quantum interference over classical conduction processes, and are fueled by  Cooperons maintaining quantum coherence. \cite{ando2002presence,takane2004quantum}

 A second case where Cooperons retain quantum coherence  is 2-D cylindrical geometries when a magnetic field is parallel to the cylinder.  In this geometry the Cooperon produces Altshuler-Aronov-Spivak oscillations where the conductance oscillates as field strength is varied.   Without  Cooperon loop repetitions, i.e. without quantum coherence, the oscillations would have a simple cosine profile.  With the loop repetitions  the Altshuler-Aronov-Spivak oscillations deviate strongly from a simple cosine signal. \cite{al1981aharonov,PhysRevB.94.205424}   
 
 These two examples prove that the absence or presence of quantum coherence in Cooperons is determined by geometry.  In 1-D systems and in 2-D cylindrical systems Cooperons can display loop repetitions.  The remaining question is how Cooperon coherence and loop repetitions are preserved in the experiments that report  linear magnetoresistance, which  generally have non-cylindrical (simply connected) 3-D or 2-D geometries.   The cause cannot be a non-Gaussian disorder distribution with large outliers, because non-Gaussian disorder penalizes higher-order processes with small weights.
 
 We speculate, very briefly, that the resolution to this question is in geometry.  It may be that in linear resistance materials transport may be locally one-dimensional, with carriers  constrained along certain race-track like trajectories.  This is  seen in snake states in graphene,  in edge states in the  quantum Hall effect and in topological insulators, and might also be realized in $C_4$ symmetry-broken (nematic) states in underdoped cuprates. \cite{PhysRevB.77.081403,PhysRevB.77.081404,PhysRevLett.107.046602,fujita2014simultaneous,PhysRevB.88.045429}  When particle motion is locked to a locally one dimensional track, at any particular point the momentum is limited to only two values differing only in sign, and therefore the Cooperon's position and momentum are locked to each other, up to the same sign.  This could allow Cooperons to retain some degree of quantum coherence and result in linear magnetoresistance.
 
 The point of introducing quantum coherence into this discussion was the hope that it could explain the extraordinarily large areas that occur in the loop area distribution of materials whose resistance is linear in field.  We close by discussing two other alternatives.  First, if spin physics and the Zeeman term were to affect conduction, then the Fourier transform of the magnetoconductance, i.e. the quantity which we call the loop area distribution, will contain information not only about loop areas but also about spin physics.   In this case the long tail of the Fourier transform may not say anything about loops but may instead be caused by some spin effect.
 
 In topological insulators the only way that spin physics can affect conduction is if conduction is ballistic instead of diffusive; if scattering is weak enough or the sample small enough that an electron can transverse the sample with very few scatterings.  It is a relatively simple matter for experimentalists to determine whether their sample is diffusive or ballistic.  Moreover, the conductance in a ballistic sample would exhibit strong dependence on the geometry, leads, and Fermi energy, and could be expected to oscillate with field. There should be little doubt about whether a particular topological insulator sample is in the diffusive regime.  In turn there should be little doubt about the validity of the loop area transform and its long tail.
 
 Second, the area detected in the loop area distribution is the sum of the areas of all loops that contribute to the material's conductance.  In non-interacting materials this sum includes only one loop.  We found that in the diffusive regime random walkers completing a single loop cannot reproduce the area distribution's logarithmic tail, which motivated our discussion of quantum coherence.  However in strongly interacting materials, where a sea of electron-hole loops determine the conductance, the story may run differently.  In this case the logarithmic distribution's fat tail and depleted head could be caused by some multi-body effect and not by quantum coherence.
 
 Most topological insulators, including the ones reported to exhibit linear magnetoresistance, are known to be weakly interacting.  \cite{zhang2012magneto,PhysRevLett.106.156808,PhysRevLett.108.266806,PhysRevB.92.081306} Therefore it seems unlikely that the long tail of their loop area distribution could be caused by interactions, and we are left again with the same conundrum that led us to quantum coherence.
 %Therefore Cooperon  coherence is protected in materials where carriers are constrained to move along locally one dimensional tracks. In conjunction with the fact that linear resistance is a kind of WAL not WL, this scenario suggests that topological physics could play an important role in   linear resistance.

%If one knows the magnetoconductance curve followed by a material, one can - without introducing any magnetic field - measure the temperature dependence of the  IR cutoff on electron loops.    
%\begin{eqnarray}
 %\mathcal{O}_i(\vec{B}=0) &=&  \int^{A_{max}} {d\vec{A}} \;   \mathcal{O}_i(\vec{A}) \\
 %\mathcal{O}_i(\vec{B}=0) &=&  \int^{A_{max}} {dA} \;   \mathcal{O}_i(A)
 %\end{eqnarray}
 %As is well known, for 2-D diffusive systems the WL contribution is $\sigma_{xx}(B=0) = \int^{A_{max}} {dA}/A \propto \ln |A_{max}|$, where $A_{max} \propto L_\phi^2 \propto L_\tau  l $ where $L_\tau  $ is the thermal decoherence length and $l$ is scattering length.  At small fields the magnetconductance is proportional to $(B A_{max})^2=(B L_\phi^2)^2$, which has allowed many transport experiments to measure the temperature dependence of $L_\phi$.  

\section{Final Comments\label{Final}}  

In our view one of the most important aspects of the present work is its methodology, which focuses on geometric analysis of electron and hole loops, and especially on the loop area distribution that can be obtained from Fourier transforms of magnetoconductance data.  We have presented a  non-perturbative framework for understanding electron and hole behavior that that can give additional physical insight.  This framework depends only on the assumption that neither the spin triplet nor the Zeeman term affect conduction, which is valid in topological insulators in the diffusive regime, and perhaps in other materials as well. We anticipate experiments focusing on the loop area distribution, with special attention to accessing large areas using carefully controlled small increments of the field, to removing leads effects, and to performing careful Fourier transforms.  We also expect increased use of vector magnets and multi-dimensional Fourier transforms.

 \appendix

\section{Alternate Regularizations \label{AlternateRegularizations}} We demonstrate in this appendix that reasonable regularizations of the Fourier transform of $x^{-1}$ produce a logarithm, and vice versa.
In our first example we perform the Fourier transform of $x^{-1}$.  
\begin{eqnarray}
\mathcal{A}(\gamma,x_0,k) & = &\int_0^{\infty} {dx} \, \cos(k x) \exp(- \gamma x ) \frac{1}{x+ x_0}  \nonumber \\
 & = &Re( \exp(x_0 (\gamma + \imath k)) \Gamma(0,x_0(\gamma +\imath k)) ) \nonumber \\
\end{eqnarray}
$\mathcal{A}(\gamma=0,x_0=1,k) $ is approximately logarithmic in the range $x = \left[0, 0.1 \right] \times 2 \pi$, as can be verified by plotting $\mathcal{A}(\gamma=0,x_0=1,k) / \log(k)$.

Our second example again performs the Fourier transform of $x^{-1}$, but with a different UV cutoff.
\begin{eqnarray}
\mathcal{B}(\gamma,x_0,k) &=& \int_0^{\infty} {dx} \, \cos(k x) \exp(- \gamma x) \frac{1}{\sqrt{1+ x^2/x_0^2}}   \nonumber \\
\end{eqnarray}
$\mathcal{B}(\gamma,x_0,k) $ can be integrated exactly, and includes Bessel, logarithmic, Struve, and hypergeometric functions.  $\mathcal{B}(\gamma=0,x_0=1,k) $ is approximately logarithmic in the range $x = \left[0, 0.1 \right] \times 2 \pi$.

Our third example performs the Fourier transform of $\log |x|$.
\begin{eqnarray}
 \mathcal{C}(\gamma,x_0,k,\nu) 
&= &\int_0^{\infty} {dx} \, \cos(k x) \exp(- \gamma x ) \log|1+ (x/x_0)^\nu |  \nonumber \\
%  \\&= Re(-2 \imath \omega^{-1} \cosh(\omega |b|) Ci(\imath \omega |b|) + \omega^{-1} \sinh(\omega |b|) (\pi + 2 Si(-\imath \omega |b|)))
\end{eqnarray}
This integral can be performed analytically for many values of  $\nu = 1, \,3/2, \, 5/3, \, 7/4, \,19/10, \, 2$, and is always proportional to $k^{-1}$ plus corrections of order $k$ at small $k$ in the range $\left[0,0.2\right]$.  For  $\nu = 3/2, \, 5/3, \, 7/4, \,19/10$ it is written in terms of Meijer functions.  The value of  $k \times \mathcal{C}(\gamma=0,x_0=1,k,\nu)$ at $k=0$ is generally a rational fraction times $\pi$. For instance, 
\begin{eqnarray}
 \mathcal{C}(\gamma,x_0,k,\nu=1) & = &Re(\omega^{-1} \exp(\omega x_0) \Gamma(0,\omega x_0) ) \nonumber \\
 \omega &= &-\imath k + \gamma .
\end{eqnarray}

%\begin{figure}
%\includegraphics[width=8.5cm]{fig1.eps} % Use any type of figure which is compatable with tex.
%\caption{(Color online) (a) ...
%}\label{fig1}
%\end{figure}

%\begin{subequations}
%\label{eq:one}
%\begin{eqnarray}
%\frac{d \theta_{c}}{dt} & = & k_{c} \theta_{p} ~ \label{eq:one_A}
%\\
%\frac{d \theta_{p}}{dt} & = & - (k_{d} + k_{c}) ~ \theta_{p}
%\label{eq:one_B}
%\end{eqnarray}
%\end{subequations}

%\begin{equation}
%\theta_{c} = \frac{k_c}{k_{+}} ~ \theta_{o} ~ (1 - e^{-k_{+} t})
%\label{eq:two}
%\end{equation}

%\begin{table}
%\caption{This is a ruled table. The table captions are
%automatically numbered.}
%\begin{ruledtabular}
%\begin{tabular}{cccc}
%One & Two & Three & Four \\
%\colrule
% 1 & 2 & 3 & 4 \\
% 2 & 4 & 6 & 8 \\
%\end{tabular}
%\end{ruledtabular}
%\label{table1}
%\end{table}

 \section{Linear-in-Temperature Resistance \label{LinearInTemperature} }

In this appendix we continue our discussion of systems whose resistance increases linearly with magnetic field.  We make remarks of a more speculative nature arguing that these systems' resistance should increase linearly with temperature. We are continuing to assume that  conduction is diffusive, with many scatterings, and that the spin triplet and the Zeeman term do not affect conduction.  We are also continuing along the lines of Section \ref{QuantumCoherence}, attributing the long $\ln A$  tail of the loop area distribution to quantum coherence. % similar to that seen in Landau levels, Cooperons in 1-D rings, and Cooperons on the surface of wires.  
Within this framework we examine the resistance's dependence on temperature.

As we have seen, the loop area distribution has a smooth logarithmic form with only two characteristic  area scales: the UV cutoff $A_0$,  and the infrared cutoff $A_{max}$   which regulates large loops.    The ultraviolet cutoff $A_0$ may be controlled by many length scales: the scattering length, the lattice spacing,  the scale of the Fermi surfaces,  and the  crossover  from WAL to WL which is controlled by the scales at which spin rotation symmetry and time reversal symmetry are broken,  etc. These mechanisms depend weakly on temperature.

The main source of temperature dependence comes instead from  the infrared cutoff $A_{max}$ which regulates large loops.  The tail of the $\ln A$ loop area distribution reflects quantum  coherence.
%harmonics where Cooperons repeat the same loop many times, resulting in total areas far larger than the area of a single unrepeated Cooperon loop.  
Regardless of which short distance physics controls the loop area distribution without quantum coherence, this physics has no control over quantum coherence, and therefore cannot determine the infrared cutoff $A_{max}$ regulating the loop area distribution.   
%The loops in the tail of the $\ln A$ distribution are far larger than those caused by pure diffusion, which indicates that $A_{max}$ is not controlled by the scattering length. 
%In particular, scattering cannot control $A_{max}$.     
 The \textit{only} mechanism available to limit the loop repetitions and thus supply a  cutoff $A_{max}$ on the loop area distribution's tail is quantum decoherence, which is controlled by temperature.

  How does the IR cutoff $A_{max}$ depend on temperature?  Since in atomic units inverse temperature has units of  area, it would be natural for $A_{max} \propto T^{-1}$ to scale inversely with temperature.  We have already noted that $A_{max} \propto T^{-1}$ does obtain in the specific case of SdH oscillations.  \cite{laikhtman1994quasiclassical} It also holds in 2-D diffusive systems when the dominant mechanism of decoherence is through electron-electron interactions. \cite{AltshulerDecoherenceLength,PhysRevB.88.155438} In linear resistance systems  $A_{max} \propto T^{-1}$ scaling should be  very robust because the loop areas in the tail are so large that $T^{-1}$ is the only area scale large enough match them.
 % both because  decoherence is the only process available to regulate the tail of $\ln A$ distribution,
 
If the decoherence area $A_{max} \propto T^{-1}$ scales inversely with temperature,  it is then a simple matter to show that at $B=0$ the resistance must be linear in temperature.  Most simply, the linear magnetoresistance is $R_{xx}(B) = \pi q_C B A_0$.  When $B=0$ the inverse area $B^{-1}$ diverges.  The only scale available to take the place of $B^{-1}$ is $T^{-1}$,  and plugging this into the magnetoresistance formula immediately gives linear in temperature resistance $R_{xx}(T) = \pi q_C T A_0$.

 Going into more detail, the logarithmic loop area distribution is $G_{xx}(A)  \approx -(\pi  A_0)^{-1} \ln |A |$. This should be regularized to give positive values because, as we have seen, in WL/WAL the Cooperon does not carry phase information and all Cooperon loop trajectories carry the same weight with the same sign.  
%  The author assumes that the loop area distribution is positive, and uses it in several places in e.g. sec III. Since these are bare fermion loops, I don’t see a justification for the distribution to be positive or even bound at all.
 One possible regularization is to add $(\pi  A_0)^{-1} \ln |A_{max} |$.  Setting $B=0$ and integrating with respect to area  gives $R_{xx}(B=0) \approx \pi   A_0/A_{max} \propto \pi A_0 T$ for $A_0 \ll A_{max} \propto T^{-1}$, plus regularization-dependent terms which are sensitive to the cutoffs.   When $A_{max} \propto T^{-1} $ is comparable to $A_0$ the linear dependence on $T$  collapses.
% This form for G_{xx}(A)  gives G \propto  ((Amax/A0) - 1)^2/2  when Amax is close to A0.
  At smaller temperatures the linear in temperature resistance is robust because $B^{-1}$ and $T^{-1}$ both act at large length scales and therefore are roughly interchangeable.   
%, and therefore   $R_{xx}(B=0) \approx  \pi    A_0 / A_{max}$ must exhibit linear-in-temperature resistance when the temperature is low enough that $A_{max} \gg A_0$.  

%in an interacting system, because $|\psi\rangle$ and $\langle \psi |$ do not scatter at the same time, conduction is sensitive to only those Cooperon loops whose time span is shorter than the characteristic fluctuation time scale of the interaction.  

% maybe discuss how at least when the B=0 intercept of the resistance is small it seems that the resistance is controlled entirely by quantum interference, like a superconductor.

 %We have shown that quantum coherence and loop repetitions lead to a resistance which scales linearly with both magnetic field and temperature as long as the quantum decoherence area $A_{max}$ is more than both the inverse field $(q_C B)^{-1}$ and  the ultraviolet cutoff $A_0$.   
 
 We summarize these mathematical arguments in three experimental observables: 
 \begin{itemize}
 \item $\alpha$ is the  coefficient of  linear temperature dependance, i.e. $R(B=0,T) = \alpha k_B T$.
 \item $\beta$ is the coefficient  of linear field dependence, i.e. $R(T=0,B)= \beta \mu_B B  $.
 \item  $\gamma = \alpha / \beta$ is the ratio of the coefficient of temperature dependence to the coefficient of field dependence.
 \end{itemize}
 These observables have the values
\begin{eqnarray}
\alpha &=& \pi A_0 / (A_{max} T) \nonumber \\
\beta & =& \pi q_C \mu_B^{-1} A_0   \nonumber \\
\gamma &= & q_C^{-1} \mu_B / (A_{max} T)
\end{eqnarray}
where $\mu_B = 1/2$  and $q_C=2$  are  the Bohr magneton and the Cooperon charge.

We can go a step further   by combining the Heisenberg uncertainty relation with a $p^2/2m$ dispersion:  $ A_{max} =  \langle (\Delta x)^2 \rangle  = \hbar^2 / \langle p^2 \rangle $, where the momentum scale is determined by  $\langle p^2 / 2 (m_C/m_e) \rangle =  T$, $m_C$ is the Cooperon mass, and $ m_e$ is the electron mass. This produces  $A_{max} T = m_e/2 m_C$.  The Heisenberg relation used here  is a quantum mechanical upper bound on the decoherence  scale which can be obtained at a given temperature.  
 
 Using this relation gives:
 \begin{eqnarray}
\alpha &=& 4 \pi  \frac{m_C}{2 m_e} A_0  = \gamma \times 4 \pi A_0 \nonumber \\
\beta & =& \pi q_C \mu_B^{-1} A_0  = 4 \pi A_0 \nonumber \\
\gamma &= & \frac{2 m_C \mu_B }{q_C m_e }= \frac{m_C}{2 m_e}
\end{eqnarray}

 These arguments indicate that $\gamma$'s physical meaning  is simply  the ratio of the Cooperon mass to that of two electrons, i.e. $\gamma = m_C/2 m_e$.  When $\gamma = 1$ the  coefficients $\alpha$ and $\beta$ of both the linear-in-temperature resistance and the linear magnetoresistance are direct measures of the UV cutoff $A_0$,  with $R_{xx}(B=0)  = \alpha T = 4 \pi A_0 T $ and $R_{xx}(T=0) = \beta \mu_B B = 2  \pi   A_0 B$.  
 %Where the carrier mass $m$ differs from the bare electron mass $m_e$, $R_{xx}(B=0) $ will be multiplied by $m/m_e$.
% The Cooperon's UV cutoff may therefore be of order $1$ in atomic units.
% Universality of the Mott-Ioffe-Regel limit in metals, 1 2 2,3 N. E. Hussey , K. Takenaka and H. Takagi

\subsection{Relation to Bad Metals}

Our derivation of linear-in-temperature resistance was made based on assumptions about diffusive scattering and the absence of spin physics.  These assumptions hold in disordered topological insulators, but in many other materials they may not be valid.  Nonetheless the linear-in-temperature resistance predicted here is also seen in  high $T_c$ superconductors (cuprates and pnictides) at temperatures above the superconducting phase.  These strongly correlated materials display a resistance which increases linearly with temperature and does not saturate.  \cite{hussey2004universality}  This  linear dependence is inconsistent both with the WAL signal which increases much more slowly with $T$ (logarithmically in 2-D and as a square root in 3-D) and with phonon based scattering which should saturate at a maximum determined by the atomic spacing.  
% WAL goes as square root of B,  and use a common formula for the dephasing cutoff on Bphi  \propto (l l_phi)^{-1} where l is the scattering length and l_phi = v_F \tau_\phi is the distance travelled before dephasing occurs and \tau_\phi \propto (k_B T)^{-1}.  This formula can vary depending on the geometry and ballistic vs. diffusive etc.
 Materials whose resistance is linear in temperature are called bad metals and are  understood to be strongly correlated, but the details of the conduction process responsible for linear resistance are not understood.  It is widely believed that linear-in-temperature resistance is correlated with high-$T_c$ superconductivity. Very recently   the linear dependence on temperature has been linked to linear magnetoresistance found in the same bad metal regime. \cite{hayes2016scaling,giraldo2018scale,kumar2018high} 
 %, which is not unnatural because both temperature and field have the same units of inverse area. 

   Most  work on linear resistance in bad metals neglects the quantum interference contributions to electronic conduction which are responsible for WL/WAL.  Linear resistance in bad metals extends up to temperatures of 1000 Kelvin, where it seems unreasonable for quantum coherence to play a role.   Most work lays the responsibility for linear resistance with a momentum relaxation time scale $\tau$ which varies inversely with temperature.  At large temperatures this time scale becomes very short, opening up a range of questions about correlation effects at short time scales  \cite{hussey2004universality}, as well as the puzzling consequence that  $\tau$   can be smaller than $\hbar / E_F$, where $E_F$ is the Fermi energy. \cite{hartnoll2015theory}   Explanations of  linear in temperature resistance have been given using resonating-valence-bond theory \cite{PhysRevLett.64.2450},  marginal Fermi liquids  \cite{PhysRevLett.63.1996},  hydrodynamics of quantum critical liquids \cite{PhysRevB.89.245116}, the SYK model with effective medium theory \cite{PhysRevX.8.021049}, %dynamical mean field theory \cite{PhysRevLett.110.086401}, 
   and bounds on diffusion in incoherent metals \cite{hartnoll2015theory}.     In Refs. \cite{PhysRevLett.64.2450,PhysRevLett.63.1996,PhysRevB.89.245116} the momentum relaxation time  $\tau \propto T^{-1}$ scales inversely with temperature, and therefore simple formulas such as the Drude formula  which have $\rho \propto \tau^{-1}$ produce  linear in temperature resistance.  Quantum interference is omitted from these simple estimates of the resistance. Ref. \cite{hartnoll2015theory} follows a similar argument but instead of $\tau$ uses a time scale $\tau_{eff} \propto T^{-1}$ which controls diffusion.    Ref.  \cite{PhysRevX.8.021049}  computes the resistance from the leading order term  in Boltzmann theory,  with a self-energy $\Sigma \propto \tau^{-1}$ which is linear in $T$. 
      %  PhysRevX.8.021049 keeps only self-energy corrections not cooperons and calls this a relaxation-time-like approximation" and obtains $\Sigma \propto T$.  
   %\cite{PhysRevLett.110.086401} (DMFT) uses Boltzmann again discusses and uses spectral density and self-energy obtained with DMFT.  The self-energy is sometimes linear but its behavior is complicated.
   The one thing in common between these works and the results presented in the present paper is that the source of linear resistance is not attributed to quantities at the atomic length scale, such as the band structure.

We list some arguments for not using the present paper's calculations as an explanation of linear resistance in bad metals:
\begin{itemize}
\item  Many believe that scattering in bad metals is so strong, and the temperatures of up to 1000 Kelvin are so large, that linear resistance cannot rely on quantum interference or quantum coherence.  Instead bad metals are in a regime of   incoherent, short lived excitations, and strong dephasing. 
\item Many believe that bad metals are not Fermi liquids; that their carriers and excitations are not fundamental (bare) electrons and holes. Nor can they  be (adiabatically) mapped to bare fermions.  Therefore it would be surprising to many if bare electrons and holes and their pairs (Cooperons) were to play a determining role in conduction.  
\item Given this doubt, it is difficult to distinguish between spin-singlet and spin-triplet components of the Cooperon, or to use reasoning about the singlet and triplet to decide whether spin has a role in  conduction. 
\item The spin physics of bad metals may be extremely rich, both in experimental signatures such as magnon spectra, and in the many theories of bad metals and high-$T_c$ superconductivity.  It is difficult to exclude the possibility that spin has a role in conduction in bad metals.
\item The theory of diffusons and Cooperons, of weak localization and weak antilocalization, generally assumes that the scattering centers are static and that scattering is elastic; it conserves energy. Moreover the diffusive conduction discussed in the present paper requires a substantial density of scatterers.  In contrast, static disorder does not seem to be a salient feature of bad metals.  Instead inelastic scattering caused by interactions is the key.
\end{itemize}

In the face of these arguments it is necessary to review the assumptions that went into the present paper.  Certainly scattering is strong in bad metals, putting them in the diffusive regime - though the scatterers are dynamic not static.  The real problem with this paper's assumptions is the requirement that spin physics does not play a role in conduction.  Because of this difficulty, we can not be sure of the transformation from field to loop area distribution; we can not be sure that the loop areas in bad metals follow a $\ln A$ distribution.

  If we were able to exclude spin effects on conduction, then the arguments of this paper would proceed  inexorably, with all the rigor of charge conservation, gauge invariance, and quantum field theory, and arrive at the $\ln A$ loop area distribution.  At this point we would have to confront the difficulty that the $A$ figuring in the $\ln A$ is the sum of areas of a vast number of strongly interacting electron-hole pairs; it is not the area of a single non-interacting random walker.  Therefore perhaps the $\ln A$ loop distribution is caused by strongly interacting physics, not by simple random walks.   This difficulty in interpreting the $\ln A$ distribution may complicate  the relation between temperature and the IR cutoff, and thus impede our arrival at linear in temperature resistance.  On the other hand $A_{max} \propto T^{-1}$ is dimensionally correct, and a large area scale like $T^{-1}$ is necessary to explore the tail of the $\ln A$ distribution.  Therefore we may arrive at linear-in-temperature resistance, even if the $\ln A$ loop area distribution is caused by strong interactions and not Cooperons.
\subsection{Experimental Results in Bad Metals}

Bearing in mind the difficulties discussed in the previous paragraphs, we comment on a few interesting experiments.  First, 
Hayes et al measured $\gamma = \alpha/ \beta$, where $\alpha$ is the coefficient of linear-in-temperature dependence and $\beta$ is the coefficient of linear-in-field dependence. \cite{hayes2016scaling}  $\gamma$ was measured in the high-$T_c$ pnictide superconductor BaFe$_2$(As$_{1-x}$P$_x$)$_{2}$ at dopings ranging between $0.31$ and $0.41$.  It was found to be  identical to one in atomic units, within the experimental error bar of $7\%$.  
% Our formulas are:
%!delta R(B,T) = q_C pi B A_0 at 1/A_{max} < q_C B
%! R(T,B=0) = pi A_0 / A_max(T) 
%! A_max = 1/ 2 T;  ! R(B=0) =  2 pi  A_0 T 
%! alpha = dR/dT = 2  pi  A_0
%! beta = mu_B^-1 dR / dB = mu_B^-1  q_C pi A_0
%! gamma = alpha / beta = 2 mu_B / q_C  = 1/2 
%-----------
% Hayes compared alpha k_B T = alpha T to beta mu_B  B. mu_B = e hbar / 2 m_e, and is 1/2 in a.u. mu_0 can be ignored because it is a symptom of their units.
% They state that k_B * 140K = 12.1 meV so 1K = (12.1/140) meV = 0.0864 meV, and mu_B * 50T = 2.89 meV so 1 T = (2.89/50) meV = 0.0578 meV.  Compare to atomic units: 3.158 * 10^5 K = 1 Hartree = 2.721 * 10^4 meV so 1 K = 0.0862 meV, and 2.35 * 10^5 T = 1 Hartree = 2.721 * 10^4 meV so 1 T = 0.116 meV.  This confirms that they were multiplying B by mu_B = 1/2 in atomic units.
% Hayes' gamma is the same as our beta.  They measure gamma^{-1} = beta/alpha = 1 for their data, 1.5 for Yb_1-x La_x Rh_2 Si2, and 0.75 for La_2-x Sr_x Cu O_4.
Further studies of the cuprate La$_{2-x}$Sr$_{x}$CuO$_4$ and of Yb$_{1-x}$La$_{x}$Rh$_2$Si$_2$ at various dopings near optimal doping have found constants between $0.7$ and $2.3$.    \cite{hayes2016scaling,giraldo2017scale}
% The 2.3 comes from Boebinger's article, which saw alpha (temperature) = 11.8 micro Ohm cm / meV divided by beta (magnetic field) = 5.2 micro Ohm cm / meV = 2.27.  In his talk Boebinger has on a slide about his data that he sees ratios between 1.5 and 2.2.  The 0.67 comes from Hayes' analysis of Yb_1-x La_x Rh_2 Si2.
 We conclude that $\gamma$ does depend on the host material, but only mildly, and that it is of order $1$.   %A very recent article examined La$_{2-x}$Sr$_{x}$CuO$_4$'s behavior at small and large $B,T$ and at several dopings between $0.19$ and $0.161$. They find a constant of proportionality between $T $ and $B$, i.e. alpha/beta, which is about $2.3$ at $x=0.19$  and decreases mildly as doping is reduced.  % it would be nice to fit the $  \rho(B) curves to find $A_{max}$ and see how that depends on temperature. 
    % Scale-invariant magnetoresistance in a cuprate superconductor
If our formula  $\gamma = m_C/2 m_e$ is correct, then   experimental observation of $\gamma=1$, i.e. a  $p^2/2m_e$ dispersion,   implies that in the  compounds studied by Refs. \cite{hayes2016scaling,giraldo2017scale} the carriers responsible for linear resistance have mass $m_C = 2 m_e$ equal to twice the bare electron mass $m_e$, and are insensitive to mass renormalization via band structure and other effects. % the ionic potential of their host material.  

We examine the   UV cutoff $A_0$'s scaling in several cuprates analyzed by  Ref. \cite{barivsic2013universal}, which determined the linear coefficient $\alpha$ in  La$_{2-x}$Sr$_{x}$CuO$_4$ (LSCO), YBa$_2$Cu$_3$O$_{6+\delta}$ (YBCO), Tl$_2$Ba$_2$CuO$_{6+\delta}$ (Tl2201), and HgBa$_2$CuO$_{4+\delta}$ (Hg1201).  Ref. \cite{barivsic2013universal} found that $\alpha$ - the coefficient of linear-in-temperature resistance -  is the same in all four compounds if one uses the sheet resistance per CuO$_2$ plane, not per unit cell.   They also found that $\alpha$ is inversely proportional to the doping $p$ for $p \leq 0.20$, where $p$ is the number of holes per unit cell, not per CuO$_2$ plane. After conversion to atomic units one finds that $\alpha =  \pi  \times 64 \, a_0^2 \times p^{-1}$.  \footnote{Up to the multiplicative constant, this formula $R \propto T/p$ is the same as Ref. \cite{PhysRevLett.64.2450}'s RVB result.  See also Ref. \cite{leggett2006quantum} for discussion of the resistance per CuO$_2$ plane.}

We surmise on dimensional grounds that $\alpha$ should not depend on the doping, but instead on the  2-D carrier density $\rho_{2D}$.   Using $p =\rho_{2D} \times \mathcal{A}$ where  $\mathcal{A}  \approx 53 a_0^2 $ is the cross-section of the unit cell in the cuprates' copper oxide plane,  we arrive at $R_{xx}(B=0) =  0.30 \times 4\pi \times  T \times \rho_{2D}^{-1}$.
% we arrive at $R_{xx}(B=0) =  0.30 \times \pi \times 4 T \times \rho_{2D}\; A_{max})^{-1} $  and $R_{xx}(B) =  0.30 \times \pi \times 4  \mu_B \times \rho_{2D}^{-1} B$, where $A_{max} = (4 T)^{-1}$.   .
% $  and $R_{xx}(B) =  0.30 \times \pi \times 4  \mu_B \times \rho_{2D}^{-1} B$, where $A_{max} = (4 T)^{-1}$.   
% Leggett has a \approx 3.84 Angstroms in his lecture notes, so A = 3.84^2 /0.529^2 = 53a_0^2

Comparison to our formula where $R_{xx}(B=0,T) = 4 \pi \gamma \times A_0 T$ suggests that in these compounds 
 the UV cutoff $A_0$ of the loop area distribution is the inverse of the carrier density, i.e. $A_0 = \rho_{2D}^{-1}$.  \footnote{Ref. \cite{hu2017universal}  has argued that  the linear coefficient of the resistivity  $dR/dT$ is proportional to $ \lambda_L^2$ across a range of cuprate, pnictide, and heavy fermion materials, where $\lambda_L$ is the London penetration depth. Since both Ref. \cite{hu2017universal}  and Ref. \cite{barivsic2013universal} analyze the same data on LSCO, this suggests a scaling relationship between   $ \lambda_L^2$ and the inverse carrier density $\rho_{2D}^{-1}$.}
This offers the possibility of determining the charge carrier density  directly from either the linear magnetoresistance or the linear in temperature resistance, without any speculation about the compound's chemistry or band structure.   

\begin{acknowledgments}
We gratefully acknowledge formative and stimulating discussions with S. Kettemann, A. Leggett, Y. Li, C. Lin, X. Dai, V. Dobrosavljevic,  H.-J. Lee, Q. Wu, T. Ohtsuki, J. Zaanen,    K. Schalm, T. Takimoto, K.-S. Kim, X. Wan, P. Niklowitz,  A. Ho, J. Saunders, L. Levitin, J. Koelzer, T. Schapers, H. Luth, C. Weyrich,  P. Hasnip, A. Kim, M. Ma, P. Coleman, S. Hayden, I. Bozovic, P. Abbamonte, and G. Parisi.  We also thank   especially A. Petrovic who discussed the manuscript during preparation.  \textbf{Funding:} We acknowledge  support from EPSRC grant EP/M011038/1. 
\end{acknowledgments}

% \pnasbreak splits and balances the columns before the references.
% If you see unexpected formatting errors, try commenting out this line
% as it can run into problems with floats and footnotes on the final page.
%\pnasbreak

% Bibliography
\bibliography{vincent}

\end{document}